\documentstyle[11pt,aaspp4,flushrt,graphicx]{article}
\def\etal{\it et al. \rm }
\begin{document}

\title{Cluster Populations in Abell 2125 and 2218}

\author{Karl Rakos}
\affil{Institute for Astronomy, University of Vienna, A-1180, Wien, Austria;
karl.rakos@chello.at}

\author{James Schombert}
\affil{Department of Physics, University of Oregon, Eugene, OR 97403;
js@abyss.uoregon.edu}

\begin{abstract}

We combine new narrow band photometry with archival WFPC2 data for A2218 ($z$=0.18) and A2125
($z$=0.25), two clusters with intermediate redshifts but very different cluster properties, in
order to examine the evolution of galaxy populations.  A2218 is a dense, elliptical-rich
cluster (Bautz-Morgan type II) similar to Coma in its evolutionary appearance; whereas, A2125
is a less dense, more open cluster (Bautz-Morgan type II-III), although similar in richness to
A2218.  The color-magnitude relation indicates that A2125 has a more developed blue population
than A2218 (the Butcher-Oemler effect), although both clusters have significant numbers of blue
galaxies (ranging in star formation rates from normal, star-forming disks to starburst systems)
as compared to a present-day cluster.  The colors of the red populations are identical in A2125
and A2218 and well fit by passive evolution models.  We are able, for the first time, to
combine archived WFPC2 images with our narrow band photometry for a color, morphological and
structural analysis of the blue Butcher-Oemler population.  We find the blue population to
composed of two sub-populations, a bright, spiral population and a fainter, dwarf starburst
population.  A2125 is richer in bright starburst systems, apparently induced by the cluster's
younger dynamical state.  In addition, a majority of the S0 population in A2125 and A2218 is
composed of bulge+disk systems, whereas, nearby clusters such as Coma are composed primarily of
lenticulars (pure disk S0's).  The structural parameters of the S0 bulges in A2125/A2218 are
identical to the structure parameters of cluster ellipticals suggesting that the disks of some
S0's in intermediate redshift clusters are stripped away with the leftover bulges evolving into
present-day ellipticals.

\end{abstract}


\section{INTRODUCTION}

Photometry of distant cluster populations can be used to explore many problems in the area of
galaxy evolution (see recent review by Dressler 2003).  Topics including color evolution, the
Butcher-Oemler effect, cD galaxy construction and S0 formation have all been
investigated through detailed photometry of galaxies in rich clusters.  However, in recent
years, cluster studies using spectroscopy have gained prominence due to the growth in
multi-fiber instruments.  This has added line indices measurements and kinematic information to
that available as new inputs to our growing community of galaxy modelers (van Dokkum \etal
2001), but photometry has the advantage of deeper limiting magnitude and spatial resolution
while still maintaining spectrophotometric quality with properly selected filter sets.

For the past decade, our team has applied a set of narrow band filters, centered on the
4000\AA\ break (a modified Str\"omgren system), to study the behavior of the near-blue
continuum in galaxies over a range of environments and redshifts.  Our technique varys from
other photometric studies in that, not only are the filters narrower and focused on continuum
regions of a galaxy's spectral distribution, but the filters are also `redshifted' to each
cluster's velocity to avoid k-corrections and provide for photometric membership selection.
The interpretation of narrow band colors differs from spectral line work in that they can
determine the mean star formation rates averaged over the past few Gyrs rather than
instantaneous rates as given by emission lines.  Thus, we find that continuum studies can
compliment the work of spectral studies, particularly with respect to the star formation
history of galaxies.

The Str\"omgren system used herein consists of four filters ($uz$,$vz$,$bz$,$yz$) which are
centered around the 4000\AA\ break.  The advantages to the Str\"omgren system for galaxy
evolution are: 1) cluster membership is assigned based on photometric criteria, thus the
ability to study lower luminosity galaxies in distant clusters which are too faint for
spectroscopy, 2) the mean star formation rate is defined by spectrophotometric criteria, and 3)
for quiescent objects (i.e. ellipticals and S0's), our color indices can resolve age and
metallicity effects (Rakos \& Schombert 2005).  Through principal component analysis
(Steindling, Brosch \& Rakos 2001), we can divide our sample by color into E (old), S-
(passive, S0-type), S (disk star formation rates) and S+ (starburst or unusually high star
formation rates).  These spectrophotometric classifications are used to identify and study the
blue population in clusters (the Butcher-Oemler effect), while the remaining red population can
be examined for metallicity and age effects.

In our past studies, we have observed the blue and red populations in a handful of intermediate
redshift clusters.  Our goal has been to study 1) the color-magnitude (mass-metallicity)
relation for the red population and mean age of ellipticals as a function of luminosity (mass),
2) identify the blue population and determine its photometric properties, 3) examine the
distribution of the red and blue populations in the clusters environment, and 4) compare the
changes in cluster populations as a function of cluster type and redshift.  In this work, now
armed with a larger set of analysis tools, we compare two clusters (A2125 and A2218) which have
similar lookback times (about 2 Gyrs), yet contrasting cluster dynamical states.  In addition,
both clusters have deep HST WFPC2 imaging allowing us to compare colors with morphological and
structural information.  This unique dataset of combined ground-based narrow band colors and
space-based imaging will allow us, for the first time, to compare color and morphological
evolution in two different cluster environments.  This paper is organized into nine sections; a
description of the observations (both ground and space-based), spectrophotometric
classification, morphological classification, structural parameters, multi-color diagrams, the
color-magnitude relation, color evolution of the red population, the blue fraction and the
characteristics of the blue population.  Throughout this work we have assumed a Benchmark
cosmology ($\Omega_M=0.3$, $\Omega_{\Lambda}=0.7$, $\Omega_k=0.0$, $H_o=75$).

\section{OBSERVATIONS}

The photometry for this project was obtained on the KPNO 4m over two nights (26/27 Jun 2001).
The instrument used was T2KB at prime focus which resulted in a plate scale of 0.48 arcsecs
per pixel.  The field of view was 14.3 arcmins square which corresponds to 2.4 Mpc for A2218
and 3.1 Mpc for A2125.  Both nights were clear with moderate seeing (0.8 arcsecs).  Each
cluster was exposed through four dithered frames of 600 secs each for a total of 2,400 secs
per filter.  Reduction used overscan bias subtraction and dome flat fields.  The flatten frames
displayed a small gradient across the field of view (on the level of 4\%), however the
flatfields removed all small features and photometry over small regions of a few tens of
arcsecs were not affected by this problem.  Calibration was obtained through spectrophotometric
standards taken through the night.  

Object selection was based on the clear identification of sources in all four filters ($uz$,
$vz$, $bz$, $yz$, see \S 3.1) at the 3$\sigma$ level.  Atmospheric absorption plus CCD
sensitivity meant that detection in the bluest $uz$ filter was the defining factor for
inclusion in the final sample.  Over 80\% of the sample is brighter than $yz=21$ (5500\AA) with the
faintest objects having magnitudes of $yz=22$.  A total of 153 galaxies were identified in all
four filters for A2125 and 277 galaxies in A2218.  The sample is complete to $yz=21$ mag
(corresponding to $-$18.4 in A2218 and $-$19.1 in A2125 using the Benchmark Model cosmology).
Typical errors at the completeness limit were 0.11 in $uz-yz$ and 0.05 in $vz-yz$ and $bz-yz$.
All the raw data is available at the author's narrow band website
(http://abyss.uoregon.edu/$\sim$js).

In addition to our ground based photometry, both A2125 and A2218 have been imaged by the Hubble
Space Telescope WFPC2 system.  A2218 has recent special attention by the HST imaging teams
since it is a well known gravitational lens cluster (Kneib \etal 1996) with five WFPC2 fields
surrounding the central core.  Each field was imaged at F606W (approximately rest frame $V$)
in 12 snaps for a total of 8,400 secs.  A2125 has one set of WFPC2 images centered on the
brightest cluster member, 2,600 secs in F606W and 2,600 secs in F814W.  In total, there are 40
objects in the A2125 field with narrow band photometry in all four filters, and 132 objects in
the A2218 fields.

The WFPC2 imaging provides a rare opportunity to test our galaxy/star rejection criteria and
our photometric redshift reliability.  As described in our previous papers, the filter set is
`redshifted' so as to avoid k-corrections by providing colors in the rest frame of the cluster
to be studied.  The methodology of rest frame Str\"omgren photometry is explored in great
detail in Steindling, Brosch \& Rakos (2001).  However, other than a comparison to cluster
members with known redshifts, we have been unable to directly assess the reliability of our
cluster membership criteria.  The WFPC2 images will allow us to 1) determine if the object is
stellar (i.e. foreground) and 2) estimate if the object is a background galaxy based on its
morphological appearance (probable foreground galaxies are obvious unless near the cluster
redshift).

There are 172 objects in the WFPC2 fields that were detected in all four filters.  Of this
sample, 29 were rejected for photometric reasons (meaning they failed the various principal
component tests to determine cluster membership by photometric redshifts, see Steindling,
Brosch \& Rakos 2001).  Inspection of the WFPC2 images finds that ten of these rejections are
clearly foreground stars.  Another 13 are background galaxies based on their small sizes,
although we note that 1/2 of these objects are elliptical and appear to be background galaxies
due to their high central surface brightnesses compared to their size (i.e. they are not dwarf
galaxies).  One galaxy is a foreground irregular.  The remaining five have peculiar colors
and/or morphologies which make it impossible to determine if they are cluster members or not.
The 143 accepted objects all had morphologies that are consistent with cluster membership.
There is some concern that the five peculiar objects that were rejected are cluster members and
would represent a unusual cluster population for this redshift.  However, it is not clear if
their peculiar colors are intrinsic or due to local problems in the photometry.  Repeat
observations would be required to resolve this question.

An extensive redshift survey was completed for A2218 (Ziegler \etal 2001). Of the 82 redshifts,
five were clearly non-cluster members (four background at $z=$1.033, 0.702, 0.699 and 0.291, one
foreground at 0.1032).  The background object at 1.033 was not detected in all four filters
(i.e. UV dropout) and rejected.  The foreground and two other background objects were correctly
rejected for anomalous colors.  The remaining object, at a redshift of 0.291, was just inside
the photometric criteria, as its redshift was just outside the cluster membership limit.  None
of the cluster galaxies (77 galaxies) were rejected by our photometric criteria.  Further
inspection of the WFPC2 images indicates that rejected objects (without known redshifts) have
the characteristics of background objects (i.e. small angular size and high surface
brightness).

Given the information gleaned from the WFPC2 and redshift data, we believe the photometric
membership criteria we have used for 20 years to be sound.  High resolution WFPC2 images
demonstrate that our redshifted filters correctly identify stars and eliminate them from the
sample.  The redshift data, although limited, indicates our procedures correctly eliminate
extreme foreground and background objects, but become more problematic for galaxies with
redshifts near the cluster redshifts or objects with extreme non-thermal colors (see
Steindling, Brosch \& Rakos 2001 for a larger discussion of this issue).  The reliability of
our photometric membership criteria lies somewhere between 99\% (the one incorrect background
galaxy by redshift) and 97\% (assuming that all five peculiar galaxies are indeed cluster
members).

\section{DISCUSSION}

\subsection{Spectrophotometric Classification}

The filter system used in our distant cluster studies (Rakos \& Schombert 1995) is a modified
Str\"omgren ($uvby$) system, modified in the sense that the filters are slightly narrower and
the $u$ filter is slightly shifted in its central wavelength as compared to the original
system.  The system we use herein is called the $uz,vz,bz,yz$ system to differentiate it from
the original $uvby$ system since our filters are specific to the rest frame of the cluster that
is being studied.  The $uz,vz,bz,yz$ system covers three regions in the near-UV and blue
portion of the spectrum that make it a powerful tool for the investigation of stellar
populations in SSP's (simple stellar population), such as star clusters, or composite systems,
such as galaxies. The first region is longward of 4600\AA, where in the influence of absorption
lines is small.  This is characteristic of the $bz$ and $yz$ filters ($\lambda_{eff}$ =
4675\AA\ and 5500\AA), which produce a temperature color index, $bz-yz$.  The second region is
a band shortward of 4600\AA, but above the Balmer discontinuity. This region is strongly
influenced by metal absorption lines (i.e. Fe, CN) particularly for spectral classes F to M
which dominate the contribution of light in old stellar populations.  This region is exploited
by the $vz$ filter ($\lambda_{eff} = 4100$\AA).  The third region is a band shortward of the
Balmer discontinuity or below the effective limit of crowding of the Balmer absorption lines.
This region is explored by the $uz$ filter ($\lambda_{eff} = 3500$\AA).  All the filters are
sufficiently narrow (FWHM = 200\AA) to sample regions of the spectrum unique to the various
physical processes of star formation and metallicity (see Rakos \etal 2001 for a fuller
description of the color system and its behavior for varying populations).

The information that can be extracted from the narrow band colors is limited by the complexity
of a galaxy's star formation history.  For example, in passive non-starforming ellipticals, the
narrow band colors can break the age-metallicity degeneracy with the caveat that the underlying
stellar population are at least 5 Gyrs old with a uniform spread in internal metallicities
(Rakos \& Schombert 2004).  Less information can be deduced from star-forming galaxies, such as
spirals, in terms of their mean ages since recent star formation dominates the older stellar
populations.  On the other hand, the filter system does provide a convenient method of
classifying galaxies based on their recent star formation rates.  To this end, a
spectrophotometric classification system using our filter system was outlined in Rakos, Maindl
\& Schombert (1996) and improved upon by the use of principal component analysis in Steindling,
Brosch \& Rakos (2001).  For our goal of investigating cluster populations, it is beneficial to
relate the photometric values to a measure of recent star formation rate rather than
morphological type, although we have the expectation that these photometric classifications
will map into morphological ones such that passive, red systems will typically be E/S0 types
and star-forming colors will typically be late-type spirals (we will examine this hypothesis
using WFPC2 images, see \S 3.2).

The newest form of our classification system divides the first principal component axis (PC1)
into four subdivisions based on mean past star formation rate; E (passive, red objects), S (star
formation rates equivalent to a normal disk galaxy), S- (transition between E and S) and S+
(starburst objects).  By comparison to SED models and nearby galaxies (Rakos, Maindl \&
Schombert 1996), the divisions along the PC1 axis are drawn such that S galaxies correspond to
those systems with spiral disk-like star formation rates (approximately 1 $M_{\sun}$ per yr)
and S+ galaxies correspond to starburst rates (approximately 10 $M_{\sun}$ per yr).  We note
that since these divisions are determined by continuum colors, versus spectral lines, they do
not represent the current star formation rate, but rather the mean star formation rate averaged
of the last few Gyrs as reflected into the optical emission by the dominant stellar population.
The red E systems display colors with no evidence of star formation in the last five Gyrs.  The
transition objects, S-, represents the fact that there is not sharp divison between the E class
and S class .  These objects display slightly bluer colors (statistically) from the passive E
class; however, the difference could be due to a recent, low-level burst of star formation or a
later epoch of galaxy formation or an extended phase of early star formation.  
In addition to classification by star formation rate, we can separate out objects with
signatures of non-thermal continuum (AGN) under the categories of A+, A and A- based on their
PC2 values colors.  It is important to remember that these classifications are based solely on
the principal components as given by the color indices from four filters.  While, in general,
these spectrophotometric classes map into morphological ones (i.e. E types are ellipticals, S-
are S0 and early-type spirals, S are late-type spirals and S+ are irregulars), this system
differs from morphology by being independent of the appearance of the galaxy and based on the
color of the dominant stellar population in a galaxy.  This is also a classification based on
integrated colors, such that large B/D galaxies (i.e. early type spirals) will generally be
found in class S- as the bulge light dominates over the disk.  This is in contrast to
classification by morphology where the existence of even a faint disk distinguishes the galaxy
from an elliptical.

The resulting population fractions based on the above photometric classifications for 153
galaxies in A2125 and 277 galaxies in A2218 are shown in Table 1.  For comparison, the
population fractions for the core of the Coma cluster are also shown in Table 1 (taken from Odell,
Schombert \& Rakos 2002).  Those objects classified as non-starforming (E and S- class)
comprise 74\% of A2125 and A2218 compared to 95\% for the core of Coma.  Star-forming and
starburst galaxies comprise the remaining 26\% compared to only 5\% in Coma.  This comparison
could be biased since the Coma sample is restricted to core regions where the
density-morphology relation predicts a deficiency of blue galaxies.  However, analyzing the
inner 500 kpc of both A2125 and A2218 (39 and 78 galaxies, listed in Table 1 as core values)
yields similar fractions.  These values are due to the well known Butcher-Oemler effect in these two
intermediate redshift clusters (Butcher \& Oemler 1978, Dressler \& Gunn 1982), the increased
fraction of star-forming and starburst objects compared to nearby clusters (see Pimbblet 2003
for a review).  We note that, the increased number of blue galaxies is clear in both these
clusters despite their different cluster morphologies, richnesses and dynamical states.

\subsection{Morphological Classification}

The availability of HST WFPC2 images for both A2125 and A2218 in the public archive provides
for a unique opportunity to compare the photometric classifications (above) with their
corresponding morphological appearance and structural properties.  The WFPC2 data for A2218 was
obtained during the 1999/2000 observing season for HST project 7343, a gravitational lens
survey.  Five fields centered on the central cD galaxy were observed in F606W for 8,400 secs in
each field.  The WFPC2 data for A2125 was more limited, consisting of a single field centered
on the brightest cluster member for 2,600 secs in F606W and F814W.  For our analysis, we have
co-added the F606W and F814W for increased S/N.  The WFPC2 fields in A2125 and A2218 contained
168 objects (30 in A2125, 138 in A2218) with matching $uvby$ photometry.  Of those, 35 were
rejected for photometric criteria, meaning identified by the PC analysis as either a foreground
star or foreground/background galaxy.

In terms of morphology, the 137 cluster members were classified by eye as either E, S0, Sa/Sb
or Sc/Irr from the WFPC2 images.  This is a simplified Hubble system as outlined by Abraham
(1999) where spirals are simply divided into large bulge spirals (Sa/Sb) or small
bulge/irregular objects.  A finer separation for spirals was deeming unnecessary given their
small numbers in rich clusters.  The separation of E's from S0's will depend on a noticeable
disk to the galaxy, the details with respect to WFPC2 images was well studied by Ellis \etal
(1997).  When the galaxy is inclined, this is detectable as a distinct pointy edge to the
isophotes versus a rounded structure (E6 in Hubble's scheme).  When the galaxy is face-on, the
distinction between E and S0 will depend on the visibility of a change in the light profile
signaled by a lens-like feature typical to S0's.  Table 2 displays the resulting matrix between
spectrophotometric classification and morphology.

The morphological distribution is similar to what is seen in other intermediate redshift
clusters (Dressler \etal 1997, Couch \etal 1998, Fasano \etal 2000) in that there is a larger
number of late-type systems in A2125/A2218 as compared to present-day clusters.  The A2125 and
A2218 WFPC2 samples contain roughly 15\% Sa to Irr type galaxies compared to 4\% for Coma.
There is also a shift in the fraction of ellipticals, approximately 25\% of the total
population in A2125 and A2218 compared to 40\% in Coma.  The percentage of S0's is slighter
higher, 57\% for A2125 and A2218 compared to 53\% in Coma.  These values are all in agreement
with morphological estimates at similar redshifts from HST or ground-based datasets (see Fasano
\etal 2000).

As expected, the red population in both clusters are predominantly early-type systems.  Of the
37 galaxies classified by morphology as elliptical, 34 (87\%) are E class photometrically, two
are S- and one is classed A+ (AGN core).  Of the 88 S0's, 56 (64\%) are E class, 15 are S- and
seven are classed as S type, where there is a statistically significant shift to bluer
narrow band colors for S0's.  Despite the fact that the photometric values are integrated, only three of
the ten large B/D systems (Sa/Sb) were classed as E or S- type, the rest were classed as S or
S+.  All the late-type systems were identified as S or S+.  Thus, the relationship between
spectrophotometric classification and morphology is surprisingly close to one-to-one.  There is
no evidence that the late-systems are `strangled', meaning a sharp decrease in star formation
as proposed by Couch \etal (2001) for A114 at $z=0.32$, since their colors are similar to
present-day spirals.  However, our colors only measure the recent star formation rate, their current
star formation rates (i.e. emission lines) are unknown.  We also note from Table 2, that the
blue population (the Butcher-Oemler population) is of a different morphological type, as well
as color, compared to present-day cluster populations and, thus, there must be color and
morphological evolution component to the Butcher-Oemler effect (Goto \etal 2003).

By morphological class, the ellipticals are the most uniform in their photometric properties
and are the reddest galaxies in the sample.  The S0 systems are similar to ellipticals in the
range of their photometric values but more often, statistically, are classed as S- or S
betraying the presence of a younger stellar population associated with their disks.  Early-type
spirals display the range of colors one would expect with a large bulge disk galaxy, and
late-type galaxies are all classed as star-forming based on their colors.  We note that the
ratio of S0's to ellipticals is approximately 2:1, similar to the ratio found in nearby
clusters (Dressler 1980), although the Coma core sample has a ratio closer to 1.

\subsection{Surface Photometry}

In addition to classification by eye, all the objects from the WFPC2 frames were subjected to a
detailed surface photometry analysis using a new automated package called ARCHANGEL (Schombert
2005).  This package takes a subimage of the area surrounding the galaxy in question,
determines local sky, cleans the subimage of nearby stars and galaxies, then fits a series of
ellipses to the 2D intensity profile.  The resulting 1D intensity profile is calibrated into a
surface brightness profile using the standard HST calibration pipeline.  The quality of surface
photometry is generally limited by two quantities, the depth of the exposure and the flatness
of the image surrounding the galaxy.  The WFPC2 frames are excellent with regard to these two
criteria, the exposures were long and the image quality was high such that the frames were
locally flat to better than 0.5\%.  The extracted surface brightness profiles had typical
errors of 0.1 mag arcsecs$^{-2}$) in galaxy envelopes (i.e. regions brighter than 24 mag
arcsecs$^{-2}$ which was due to a combination of error along the elliptical isophote, sky error
and zeropoint uncertainity.  Errors increased to 0.5 mag arcsecs$^{-2}$ at a limiting surface
brightnesses of 28 mag arcsecs$^{-2}$ where the fits were halted.  This compares well with
other surface photometry studies of cluster galaxies (Schombert 1986) and probably represents
an upper boundary for WFPC2 data quality due to limitations in flatfielding.  Comparison
surface brightness profiles for the Coma cluster was extracted from DSS images and analyzed in
an identical manner.

The resulting surface brightness profiles are then fit under the assumption that the galaxies
is either a pure exponential disk, a $r^{1/4}$ shape (bulge or elliptical) or a combination of
both (early-type disk galaxy with a bulge and disk).  Classification by structure hopefully
avoids the problem that subjective visual classification may have, especially for the E/S0
catalogories (Andreon 1998).  Misclassification of an S0 as an elliptical may occur if the disk
is small or very faint, however if a majority of the mass or size of a galaxy is $r^{1/4}$ then
its classification as elliptical seems warrented.  For the 137 galaxies in A2125 and A2218 with
HST imaging, 37 (27\%) are $r^{1/4}$ shaped, 27 (20\%) are pure exponential disks and the
remaining 73 (53\%) are bulge+disk objects (see Table 3).  This distribution by profile type
differs significantly from the distribution found in the Coma sample where 43\% of the galaxies
are $r^{1/4}$ shaped, 36\% are pure exponential disks and 20\% are bulge+disk objects.  The
difference between the Coma and A2125/A2218 samples being primarily due to the increase in
$r^{1/4}$ shaped objects (ellipticals) and the decrease in bulge+disk objects (S0's and
early-type spirals) relative to  Coma.  The blue population (S/S+) in A2125 and A2218 are
mostly disk systems (17 out of 27, the remain 10 are bulge+disk objects).

Both clusters display a ratio of S0's to ellipticals near two (under the assumption that all
red r$^{1/4}$ objects are ellipticals and all red B+D or D objects are S0's), which is similar to the
value determined for nearby clusters by Dressler (1980).  But, a HST survey of ten clusters
between $z=$0.3 and 0.6 (Dressler \etal 1997) finds the S0 to elliptical ratio to be
approximately one by redshifts of 0.3 (although we note that Fabricant, Franx \& van Dokkum
(2000) find a ratio of 1.6 for CL1358+62 at $z=0.33$ which is more in agreement with our
values).  Our S0 fractions (approximately 55\%) for A2125 and A2218 agrees well with the
measured morphological trends of 25 clusters ranging from $z=0$ to 0.6 by Fasano \etal (2000)
(see their Figure 9), but our elliptical fraction of 25\% is slightly below their values of 30\%
to 35\% at redshifts of 0.2.  This difference may be due to subtle differences in visual
morphology versus surface photometry (Andreon 1998), but the structural differences between
ellipticals and S0's were clear to visual inspection if detectable in their surface brightness
profiles.  

\begin{figure}
\centering
\includegraphics[width=9cm]{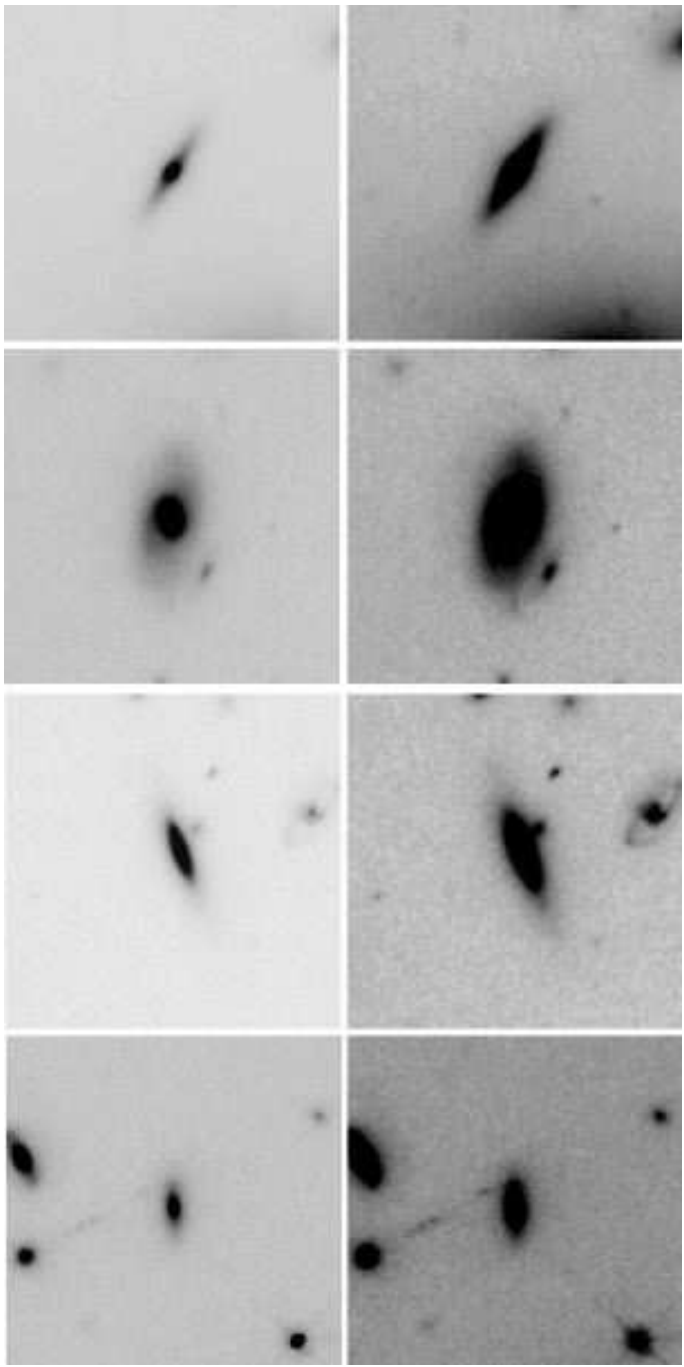}
\caption{The above figure is a mosaic of four S0 galaxies in A2218, two classical S0's and two
lenticulars.  The left hand panels are of low contrast to emphasize the core regions.  The
right hand panels are of high contrast to display the outer isophotes.  The top two galaxies
are bulge+disk systems (B+D), meaning they have distinct bulge and disk components in their
surface brightness profiles.  The bottom two galaxies are pure disk systems (lenticulars).  The
differences in their inner structure is clear in the low contrast images, while their visual
morphology is similar.  A2125 and A2218 are unusual (compared to Coma) by having a high
fraction of B+D S0's compared to lenticulars.}
\end{figure}

Comparing visual morphology to surface brightness structure (Table 4), we find that a majority
of the morphologically classified S0 galaxies in A2125 and A2218 are composed of a bulge plus
disk (B+D) structure.  In contrast, Coma S0's are mostly pure disk systems (lenticulars under
the nomenclature proposed by van den Bergh 1990).  The distinction between B+D S0's and
lenticulars (disk only) is not an artifact of the fitting process.  Figure 1 displays a mosaic
of two B+D S0's and two lenticulars from A2218.  The left panels are low contrast images to
emphasize the inner bulge regions, the right panels are high contrast images to emphasize the
outer disk.  It is clear from this Figure that B+D S0's have a separate, round bulge region
with a distinct and flattened disk.  The lenticulars have no break in their profile to signal a
bulge, nor any change in flattening of the inner isophotes.

The fact that S0 galaxies divide into two populations by structure has been previously noted
before in an analysis of the RSA (van den Bergh 1994).  In that study, a comparison is made
between the colors and luminosity functions of E's, S0's and Sa's where the S0 population is
found to be intermediate in color and gas/dust content between E's and Sa's, however, their
absolute luminosity has a lower mean value than either E's or Sa's.  Comparing the structure of
galaxies in our Coma sample, we find that the S0's in the core of Coma are also composed of a
dual population, where the brighter S0's are the B+D objects and the fainter S0's are
lenticulars.  This is the same distinction that is found in the A2125 and A2218, where the S0 galaxies
neatly divide into the bright objects (bulge+disk) and the fainter lenticulars (disk only).
However, when we compare the population fractions in A2125/A2218 to the Coma core sample, Coma
is deficient in the bright S0's and the total Coma S0 population has a mean luminosity that is
1.5 mags fainter than the combined A2125/A2218 S0 population.

\begin{figure}
\centering
\includegraphics[width=16cm]{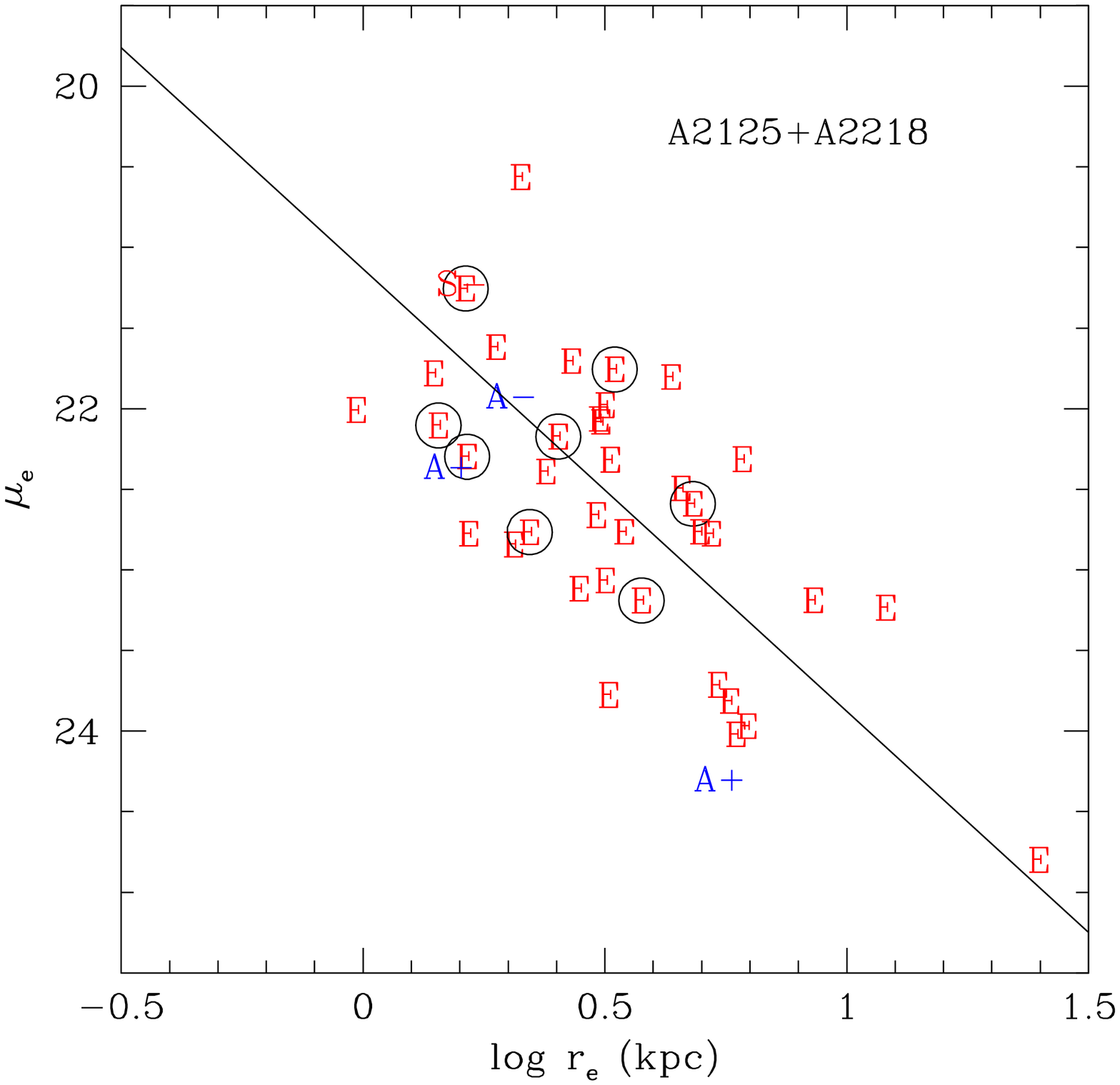}
\caption{The r$^{1/4}$ structural parameters effective radius ($r_e$) versus effective surface
brightness ($\mu_e$) for all bulges ($B/D > 0.1$) and elliptical shaped objects in the WFPC2
fields for A2125 and A2218.  Symbols display the galaxy's photometric classification.  Circled
symbols are B+D S0's.  The solid line is the relation for ellipticals from Kormendy (1980)
corrected for surfqace brightness dimming at the mean redshift of A2125 and A2218.  Galaxies in
both clusters with r$^{1/4}$ profiles (ellipticals and S0 bulges) follow the same structural
relation as present-day ellipticals.}
\end{figure}

One explanation for the change in the S0 population (from B+D dominant to lenticular dominant)
is that S0 disks fade (or are destroyed) on the short cosmological time from $z=0.2$
such that past B+D S0's now appear as a pure $r^{1/4}$ ellipticals today.  This can tested 
by examining the correlations of the structural parameters derived from the surface photometry.
Ellipticals and S0 bulges, described by an $r^{1/4}$ shape, are parameterized by the effective
surface brightness ($\mu_e$) and effective radius ($r_e$).  These two values are shown in
Figure 2 where each galaxy is represented by its photometric classification.  For clarity, only
pure r$^{1/4}$ and large bulge ($B/D > 0.1$) galaxies are shown as small bulge fits suffered
from numerical errors due to the dominance of the disk intensity.  For comparison, the solid
line is from Kormendy (1980) converted to the Benchmark cosmology and corrected for surface
brightness dimming (there are no k-corrections since F606W is equivalent to rest-frame $V$ at
these redshifts).  Despite the lookback time to A2125 and A2218, their $r^{1/4}$ structural
correlations are identical to present-day galaxies.  The scatter about the Kormendy relation is
similar to that of other bright cluster ellipticals (see Figure 6, Schombert 1986).  The
circled data points represent large S0 bulges which follow the same relationship as the
ellipticals, meaning that if these S0's were stripped of their disks, they would have the same
structural properties as present-day cluster ellipticals.

\begin{figure}
\centering
\includegraphics[width=16cm]{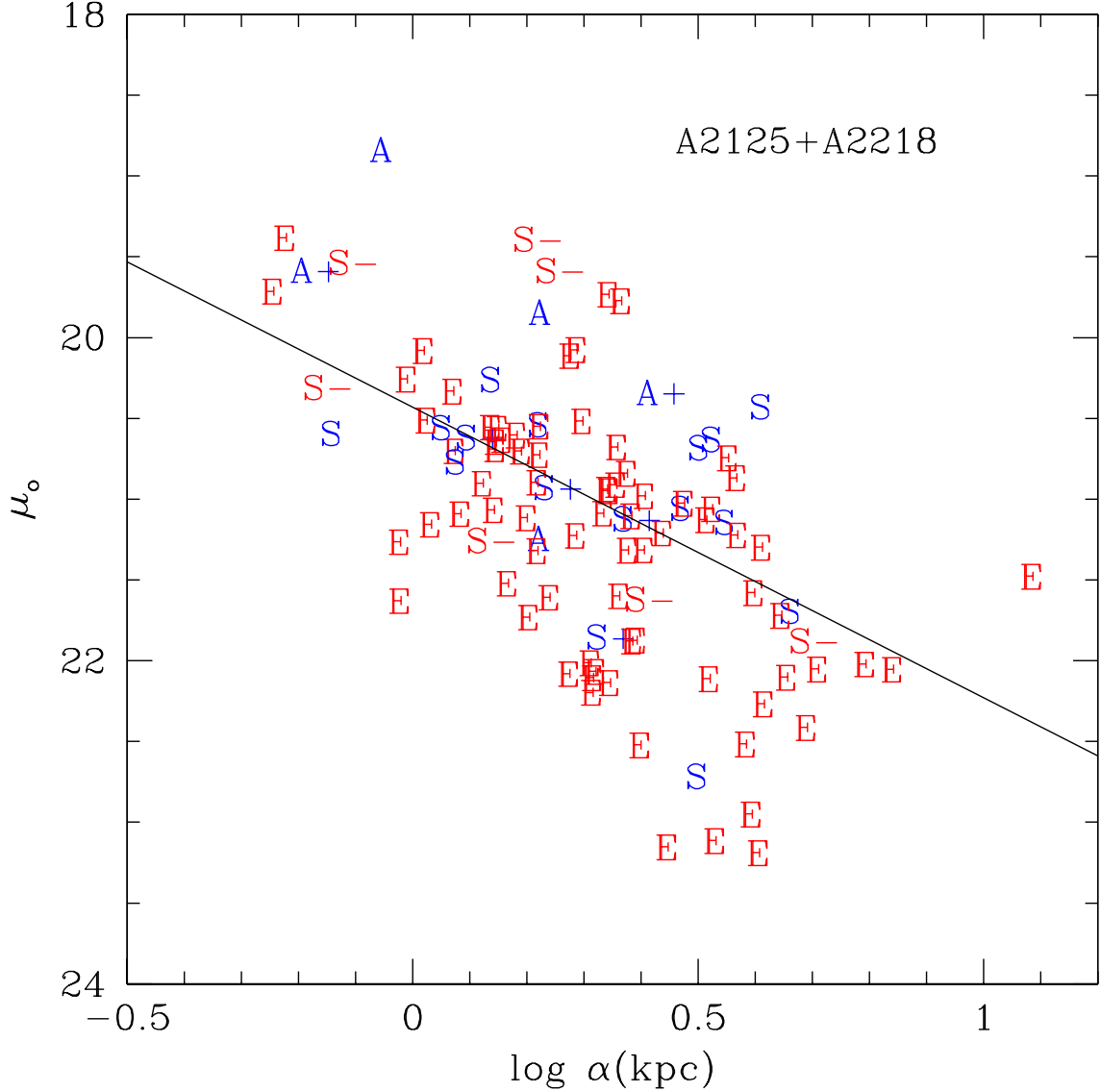}
\caption{The disk scale length ($\alpha$) versus the disk central surface brightness for all
the B+D and D type galaxies in A2125 and A2218.  Symbols display the galaxy's photometric
classification.  The solid line is from MacArthur \etal (2004) and represents the mean relation
between central surface brightness and scale length for a range of Hubble classes.  The blue
population follows the MacArthur \etal relation (higher central surface brightness due to
ongoing star formation). A majority of the systems below the solid line are S0's with faded disks.}
\end{figure}

Disk structural parameters are more difficult to interpret as there is no clear scale length to
characteristic surface brightness relation as is found for ellipticals.  This lack of
correlation between scale length and surface brightness is due to the wide range in past star
formation rates and formation processes in spirals which reflects into a range of central
surface brightnesses.  Despite varying star formation histories, there is a trend of increasing
central surface brightness for early-type spirals.  The most recent study of the disk structure
of spirals is MacArthur \etal (2004).  Taking a straight line through their brightest spirals
(see their Figure 1), produces the line shown in Figure 3 (corrected for the distance to
A2218).  Compared to this relationship, most of the blue disk galaxies in A2125 and A2218 have
similar central surface brightnesses as the typical spiral from MacArthur \etal .  On the other
hand, many of the red disks (S0's) are fainter than the MacArthur \etal relation.  This
suggests a connection between the lack of lenticular S0's in Coma in the sense that these
systems in A2125 and A2218 are faded compared to their present-day counterparts and, perhaps,
are `missing' in the Coma sample simply because they have converted into elliptical shaped
galaxies with invisible (with respect to surface brightness) disks.  In addition, if stellar
luminosity traces mass, then the mass density of these disks is low and they are susceptible to
stripping by the cluster tidal field.

\subsection{Multi-color Diagrams}

Our primary diagnostic for understanding the colors of cluster galaxies is the multi-color
diagrams $uz-yz$,$bz-yz$, $vz-yz$,$bz-yz$ and $vz-yz$,$mz$.  Figures 4 and 5 display these
multi-color diagrams for all the detected galaxies in A2125 (153 galaxies) and A2218 (277 galaxies).
All galaxies classified as E or S- are marked in red, classes S and S+ (as well as A) are
marked in blue.  The size of the symbol indicates the galaxy's absolute magnitude.  In general,
the multi-color data for A2125 and A2218 follows the trends for other intermediate redshift
rich clusters (Rakos \& Schombert 1995).  Like present-day clusters (i.e. Coma), a majority of
the galaxies have red, passive colors usually associated with elliptical and S0 morphologies.
Unlike nearby rich clusters, there are a large number of blue, star-forming or starburst
systems in each cluster (the Butcher-Oemler effect).  The red population has a relatively tight
grouping in each diagram, the correlation between various colors reflects the differences in
metallicity with galaxy mass (see below).  Each cluster's blue population is distinct from the
red population.  A2125 differs from A2218 in having a grouping of very bright blue galaxies
around $bz-yz$=0.1 and $vz-yz$=0.0 (see discussion of color-magnitude diagram below).

\begin{figure}
\centering
\includegraphics[width=16cm]{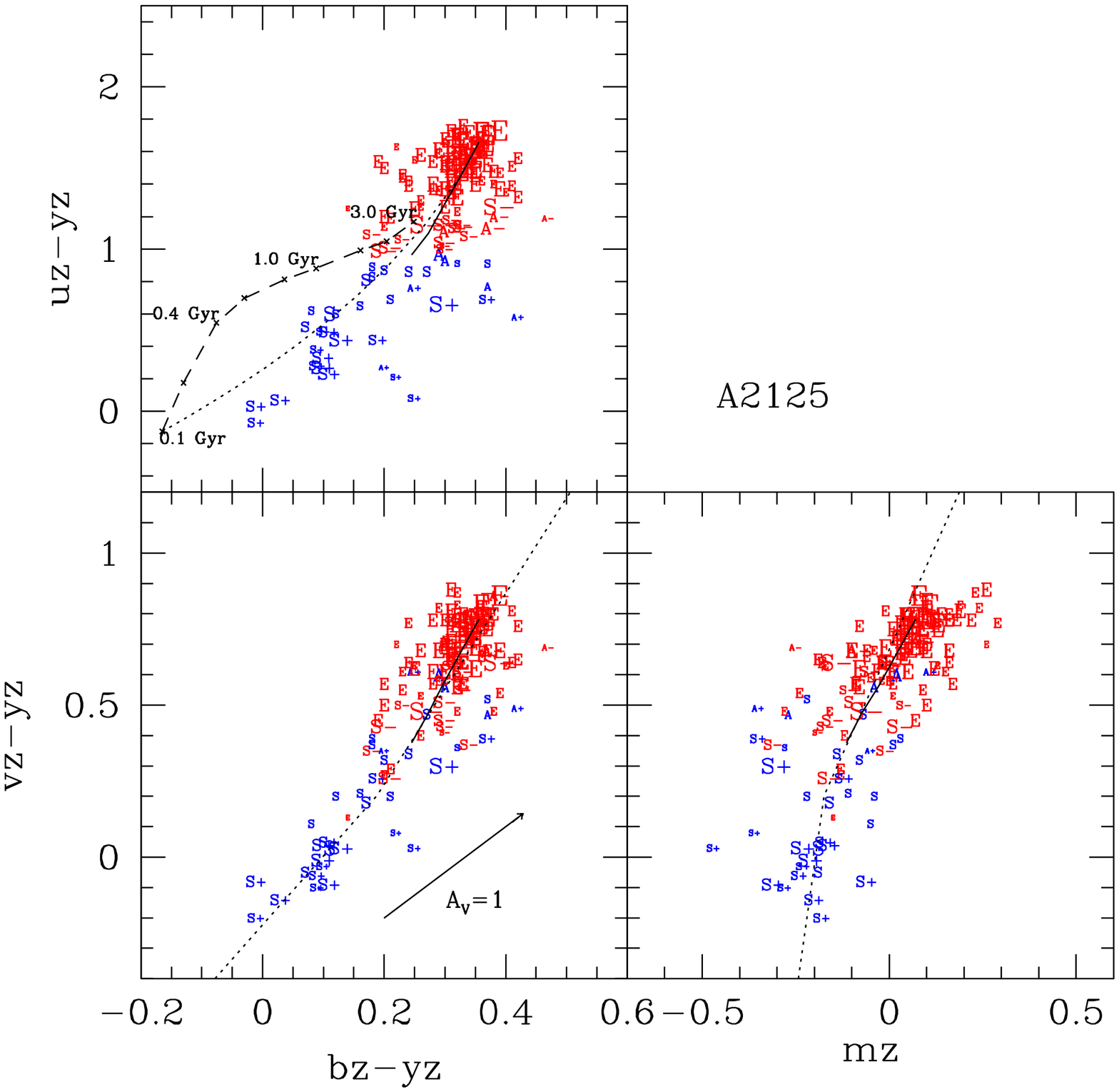}
\caption{Multi-color diagrams for A2125.   All three colors and the $mz$ index are shown for
galaxies determined to be cluster members by photometric criteria.  Photometric classification
is shown by symbol type (E, S-, S, S+) and the red population is denoted by that color as is
the blue population.  Symbol size indicates absolute luminosity of the object.  The solid line
in each diagram displays the 13 Gyrs models for metallicities ranging from $-$0.7 to +0.4.  The
dotted line in the $vz-yz$ and $mz$ diagrams represents the 99,000 galaxies from the SDSS
sample (Smolcic \etal 2004).  The dashed line in the $uz-yz$ diagram is the range in age for a
solar metallicity model.  The dotted line in the same diagram displays a `frosting' model, i.e.
adding increasing fractions of a 0.1 Gyr population to a 13 Gyrs population.  A reddening
vector is shown in the $vz-yz$ diagram.}
\end{figure}

\begin{figure}
\centering
\includegraphics[width=16cm]{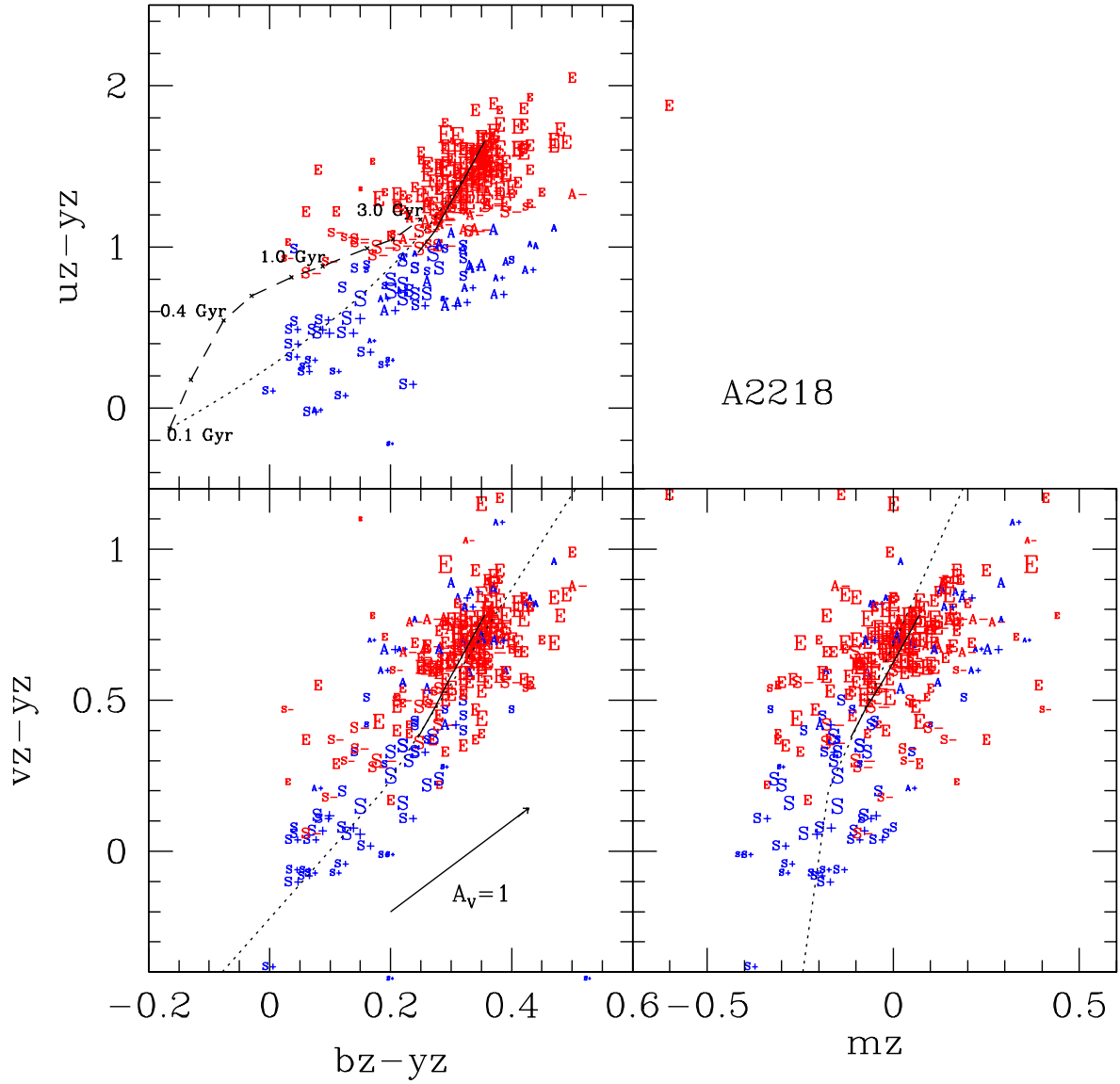}
\caption{Multi-color diagrams for A2218.  Symbols and lines are the same as Figure 4.}
\end{figure}

The dotted lines in $vz-yz$,$bz-yz$ and $vz-yz$,$mz$ diagrams are the mean color track of
99,000 SDSS galaxies (Smolcic \etal 2004).  These colors were extracted from a flux limited
sample of SDSS spectra (which does not cover the $uz$ filter), converted to rest frame.  The
scatter about this locus is 0.03 mags.  Both clusters track the same locus, although with more
scatter reflecting the lower S/N data of our sample.  Also indicated in the diagrams are a
series of SED models taken from Bruzual \& Charlot (2003) and convolved to our
multi-metallicity models (see Rakos \etal 2001).  The solid line in all three panels displays
the run of color for a 13 Gyrs population with mean metallicities ranging from Fe/H =
$-$0.8 to $+$0.4 (blue to red).  The red galaxies in our sample (classes E and S-) nicely
bracket the 13 Gyrs models.  The dashed line in the $uz-yz$ panel is a run of pure age
populations of solar metallicity ranging from 0.1 Gyr to 3 Gyrs old.  Even the bluest galaxies
in our sample fail to match a pure young population indicating the presence of an old stellar
population in all cluster galaxies.  The dotted line in the same panel displays the fractional
mixture of a 0.1 Gyr population and a 13 Gyrs population, so-called `frosting' models.  The
blue galaxy population colors are well matched to this type of model indicating that even
starburst galaxies have a significant underlying older population in their integrated colors.

Galaxies composed of a single aged and old stellar population will form a linear sequence in
this diagram with position being determined solely by the mean metallicity of the underlying
stars.  Since these are integrated colors, the metallicity reflected by the color is a
luminosity weighted value (versus spectroscopic values which are surface brightness weighted
and usually represent core values).  Based on our metallicity calibration from globular
clusters (Rakos \& Schombert 2005), the galaxies with the reddest colors have mean
metallicities just above solar, dropping to values of [Fe/H] between $-$0.5 and $-$1.0 for the
galaxies a magnitude fainter than $L_*$.  In our previous work, we have demonstrated an age
dependence for ellipticals (E class) with luminosity (Odell, Schombert \& Rakos 2002), but this
effect is minor in the $vz-yz$,$bz-yz$ diagram compared to metallicity effects.  While
star-forming galaxies appear to follow the same linear relation in Figures 4 and 5, in fact
their colors are dominated by a number of young, massive stars and not metallicity effects.
In addition, several S and S+ class objects lie below the sequence of ellipticals along the
extinction vector from their bluer counterparts indicating the prescence of dust.

The $vz-yz$,$mz$ diagram ($mz=(vz-bz)-(bz-yz)$) is used to separate systems with a history of
moderate to strong star formation from those with passive histories (ellipticals), and also to
distinguish non-thermal colors (AGN's) from stellar colors.  The photometric classification is
best done with PC analysis (which uses all the color information, see Steindling, Brosch \&
Rakos 2001), although one can see that almost all star-forming galaxies have low $mz$ values.
The information in this diagram is further confused as one samples deeper into a cluster's
luminosity function.  Faint dwarf ellipticals ($M_{5500} > -17$) have low $mz$ values due to a
combination of extreme old age and low metallicity (see Rakos \& Schombert 2003).  However,
this is not a concern for this study since we limit ourselves to galaxies with $M_{5500}$
brighter than $-$18.

The $uz-yz$,$bz-yz$ diagram is a measure of recent star formation rate.  Even objects with
non-thermal colors easily separate from passive red objects in this diagram since the amount of
UV light is directly proportional to the number of massive stars, aside from extinction
effects.  Comparison with SED models indicates that galaxies with $uz-yz$ values between 0.7
and 1.0 have mean star formation rates (over the last 0.5 Gyrs) similar to spiral disks (i.e.
1-2 $M_{\sun}$/yr, assuming near solar mean metallicities).  Galaxies with $uz-yz$ values below
0.7 are undergoing a starburst with SFR at, or greater than, 10 $M_{\sun}$/yr.  A majority of
the star-forming galaxies in both A2125 and A2218 display $uz-yz$ colors in agreement with the
`frosting models, i.e. an older population undergoing a recent burst of star formation.  The
remaining blue galaxies appear to be star-forming colors with medium extinction values ($A_V$
between 0.5 and 1.5).

\subsection{Color-Magnitude Relation}

The color-magnitude diagram for ellipticals is the most common tool for investigating red
populations in clusters and is considered to be a classic demonstration of the varying
metallicity with galaxy mass for old cluster populations.  Discovered in the days of
photoelectric photometry (Faber 1973, Visvanathan \& Sandage 1977), the color-magnitude
relation (CMR) is typically interpreted as the increasing ability for deeper gravitational
wells (higher luminosity systems) to maintain a chemically enriched ISM (Larson 1974, Arimoto
\& Yoshii 1987).  Early CMR's were observed in near-blue colors, such as $U-V$, however, since
then the CMR has been investigated from the far-UV to the near-IR (Pahre 1999, Ellis \etal
1997). With respect to the $uvby$ filter system, we have previously presented a full analysis
of our filter system's behavior with galaxy luminosity in Coma and Fornax (Odell, Schombert \&
Rakos 2002).

While the color-magnitude relation has been studied at many wavelengths (Pahre 1999) and
spectroscopically, using line indices and velocity dispersions in the place of color and
luminosity (Kuntschner 2000, Trager \etal 2000), there is no theoretical expectation that the
relationship should be a linear or tightly constrained.  In other words, the star
formation history of galaxies can be diverse and there is no reason why the CMR should exist
nor why its scatter is so small (Andreon 2003).  Mean stellar age can have a strong effect on
the CMR, but from analysis of galaxy color and line indices (Kuntschner 2000, Trager \etal
2000) and due to a lack of visible change in the CMR with redshift (Stanford, Eisenhardt \&
Dickinson 1998), it is generally accepted that the CMR reflects a mass-metallicity sequence in
an old stellar population.  The low scatter to the CMR, in what is apparently a solely
metallicity effect, is interpreted to mean that the formation of ellipticals is coeval and
uniform (the so-called monolithic model of galaxy formation, see Andreon 2003 and Ellis \etal
1997 for a review).  Hierarchical models of galaxy formation (Kauffmann \& Charlot 1998) can
reproduce the CMR, but the low scatter makes the process highly contrived.

\begin{figure}
\centering
\includegraphics[width=16cm]{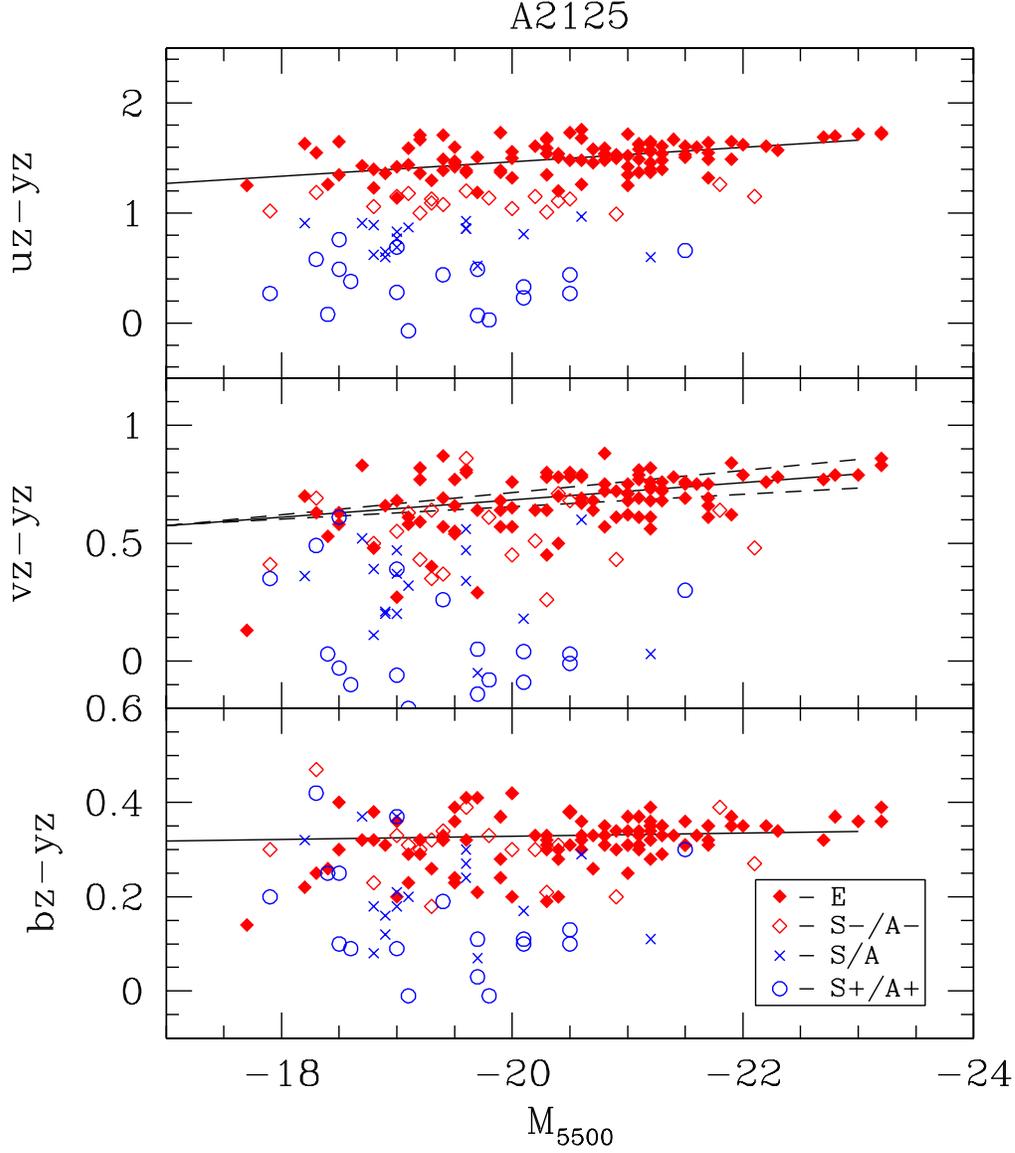}
\caption{The color-magnitude diagram in three colors for A2125.  Symbol types for the different
photometric classifications are shown.  The red population (class E and S-) are shown in red,
the blue population (S and S+) are shown in blue.  The solid line is the fit to the Coma/Fornax
data from Odell, Schombert \& Rakos (2002), dashed lines in the $vz-yz$ diagram display the
3$\sigma$ errors on the fit.  The similarity to Coma is striking, even reproducing the flat
$bz-yz$ relation which is unexpected in Coma, or this paper's data, as metallicity effects
should be detectable in the $bz-yz$ color.  A lack of slope indicates a competing age effect
such that lower mass galaxies are older.}
\end{figure}

\begin{figure}
\centering
\includegraphics[width=16cm]{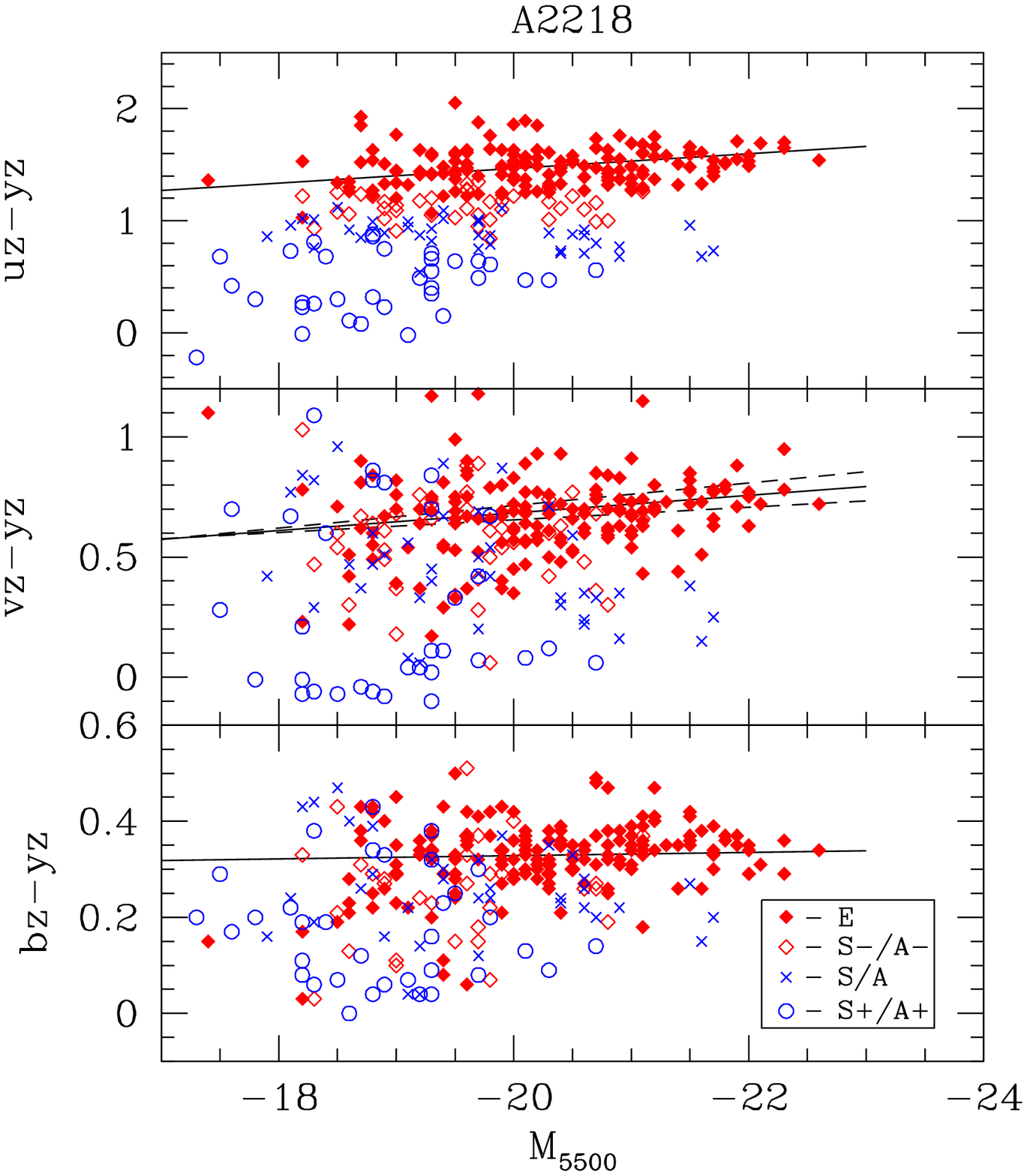}
\caption{The color-magnitude diagram in three colors for A2218.  The symbols and lines are the
same as Figure 6.}
\end{figure}

The color-magnitude relation for A2125 and A2218 is shown in Figures 6 and 7, where the top
panel displays the near-UV color, $uz-yz$, the middle panel displays the metallicity color,
$vz-yz$, and the bottom panel displays the continuum color, $bz-yz$, as a function of
$M_{5500}$.  The solid lines are fits to the color-magnitude relation for Coma with the dotted
lines in $vz-yz$ displaying the 3$\sigma$ variation to the Coma fit (note, these fits were made
only to those systems which were morphologically classified as ellipticals in Coma, see Odell,
Schombert \& Rakos 2002).  While our discussion below will be confined to the photometric E
class objects, all the cluster data from E to S+ are shown for completeness.

The $uz-yz$ CMR diagram is nearly equivalent to $U-V$ based on central filter wavelength,
although the $uz-yz$ color samples a much narrower portion of a galaxy's SED.  While the $uz$ region
of the spectrum is sensitive to metallicity effects, it is much more sensitive to light from
hot, massive stars, i.e. recent star formation and mean age.  Yet, the correlation between
luminosity and color is the strongest, and with the lowest scatter, in the $uz-yz$ colors.
Luminosity and $vz-yz$ color is also well correlated, in agreement with the expectation that
$vz-yz$ is a strong metallicity indicator.  However, the relation for $bz-yz$ is much weaker,
although in agreement with Coma $bz-yz$ data.  In all three diagrams, the scatter around the
Coma fits is larger than observational error.

If the CMR is reflecting a mass-metallicity relationship, then interpretation requires
converted luminosity to mass and color to [Fe/H].  Conversion of luminosity to mass requires a
calibration of $M/L$, which for ellipticals is narrow and linear (Bender, Burstein \& Faber
1992).  Relating the color axis to metallicity is more problematic.  Our initial calibration
of $vz-yz$ to [Fe/H] was through the use of globular cluster colors and spectroscopic [Fe/H]
values.  However, globular clusters are single metallicity plus single age systems and galaxies must
be composed of a range of stellar populations with varying metallicities and, possibly, varying
age.  While a multi-metallicity population is relatively easy to calculate from globular
cluster colors, a varying age population is not due to the lack of a range on ages in globular
cluster system around the Milky Way.  For this reason, we have adopted a system of 
globular cluster metallicities to calibrate models of simple stellar populations (SSP's, see
Schulz \etal 2002), and then using these models to explore age variation.

Our technique for calibrating the $vz-yz$ to [Fe/H] is outlined in Rakos \& Schombert (2004).
From these empirical calibrations, we extend the metallicity scale to a range of SSP's
available in the literature, then construct a model for a single galaxy as the sum of a
Lorenzian distribution of metallicities where the peak [Fe/H] value is a free parameter and a
corresponding long tail towards low [Fe/H] values.  Each metallicity bin has a corresponding
population color which is summed by luminosity weight.  Each galaxy is characteristized by a
mean metallicity ($<$Fe/H$>$), a numerical mean of the total population as weighted by
luminosity (this value is always slightly less than the peak metallicity, which is more typical
of spectroscopic values taken from the bright, metal-rich core regions of galaxies).  The
resulting multi-metallicity models explain all the details of the $<$Fe/H$>$ versus $vz-yz$
diagram determined from spectroscopic [Fe/H] values (see Rakos \etal 2001).

Armed with this calibration method, we find that $vz-yz$ colors near 0.75 correspond to
$<$Fe/H$>$ values of $+$0.2 for the high luminosity galaxies in A2125 and A2218.  Likewise,
$vz-yz$ colors of 0.60 for low luminosity galaxies corresponds to $<$Fe/H$>$ values of $-$0.5
(these axis are shown in Figure 9 to be discussed below).  Of course, it is naive to expect
that each galaxy arrives at its final metallicity based solely on mass given the numerous
environmental processes that occur in rich clusters.  However, the linear nature of the
color-magnitude (in a log-log parameter space) does imply that a coherent process is in
operation for the stellar populations in elliptical galaxies, if their point of origin extends
to redshifts greater than 5.

While it is universally accepted that a majority of the CMR is due to metallicity, age effects
can play an important component.  In fact, one of the key differences between the monolithic
scenarios of elliptical formation (Eggen \etal 1962) and the hierarchical model (Kauffmann 1996)
is the role that age will play.  Monolithic formation implies a very narrow range of age within
an ellipticals stellar population, whereas hierarchical formation could result in a narrow
range if all the component galaxies have single formation epochs or a range in age if the
merging components have varying formation epochs (Kauffmann \& Charlot 1998).

Observationally determining an age difference is a difficult task even for narrow band filters
due to the age-metallicity degeneracy.  For galaxies with old stellar populations, the two
color diagram in Figures 4 and 5 can resolve age differences between 2 to 3 Gyrs (see Rakos \&
Schombert 2004) with the caveat that the dominant stellar population in the galaxy in question
is older than 5 Gyrs and has not undergone a recent star formation event or absorbed a younger
stellar population.  A younger mean age displays itself through the continuum color, $bz-yz$.
According to expectations from our multi-metallicity models, and the observed slope of CMR for $vz-yz$,
we find there should be a shallow, but measurable slope in $bz-yz$ with luminosity.  However, analysis
of the $bz-yz$ colors in both A2125 and A2218 (and Coma plus Fornax) indicates zero slope in
$bz-yz$ versus luminosity, even though the metallicity changes a full dex for the luminosity range.  The most
obvious interpretation is a competing age effect in the direction such that lower mass galaxies
are older.  While, to first order, this appears to be a stunning victory for the hierarchical
scenarios (dwarf galaxies formed first and more massive systems are constructed from a
combination of old dwarfs and spirals), in fact, this trend could be derived in many ways.  For
example, bright ellipticals in rich clusters clearly have a history of mergers and the
cannibalism of even one star-forming spiral or irregular would push their continuum colors
bluer and force an underestimation of their mean ages.  This interpretation also draws some
support from the fact that the scatter of the brightest galaxies in the $bz-yz$ CMR is greater
than any other color, which is predicted for a random process such as mergers.

\begin{figure}
\centering
\includegraphics[width=16cm]{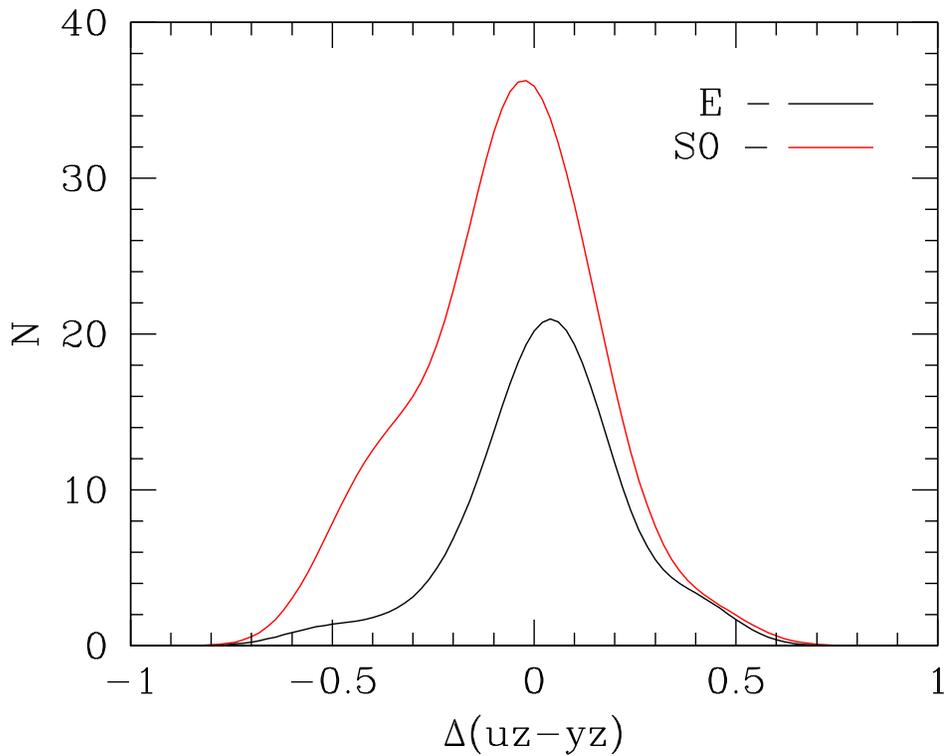}
\vspace{-4cm}
\caption{Normalized histogram of the difference in the $uz-yz$ colors as measured from the mean
CMR line for morphologically determined ellipticals and S0's.  The ellipticals in A2125 and
A2218 display a mean of zero around the CMR; whereas the S0's from the WFPC2 sample have a
clear tail to the blue.  Since $uz-yz$ measures recent star formation, when corrected for
metallicity effects of the CMR, then the blue tail indicates a younger mean age for S0's or a
recent episode of star formation.  A shift of 0.2 in $uz-yz$ color occurs between a 5 Gyrs and
13 Gyrs population. Thus, either S0's had their last star formation 5 Gyrs ago, or have
experienced a more recent episode involving some small fraction of the total stellar population
on top of a 13 Gyrs population.}
\end{figure}

The morphological information from the WFPC2 images allows us to inspect the CMR for
differences between ellipticals and S0's since the claim is made, based on spectroscopic
measurements of Coma galaxies, that a significant fraction of S0's have undergone star
formation in the last 5 Gyrs (Poggianti \etal 2001).  This is a key statement concerning the
star formation history of galaxies based solely on their morphology appearance and, in this case,
the existence of a disk implies an extended phase of star formation.  There are 112 galaxies in
the WFPC2 fields of A2125 and A2218 with morphological classifications of E (36 galaxies) or S0
(76 galaxies).  A comparison of their colors over a range of luminosities will be biased due to
the CMR.  To avoid this effect, we have examined a histogram of the differences from the $uz-yz$ CMR
for each galaxy type.  Figure 8 displays the resulting histogram and it is clear that the
ellipticals form a well shaped gaussian around the zeropoint, but that S0's have a slightly
different mean value and a long blue tail.  This confirms the Smail \etal (2001) result that
S0's display slightly bluer integrated colors (in a range of optical and near-IR colors)
consistent with a prolonged period of star formation as compared to ellipticals (for a
dissenting view, see Jones, Smail \& Couch 2000).  We note that Ziegler \etal (2001) find no
difference in age between the ellipticals and S0's in A2218, although theirs was a
spectroscopic study (i.e. galaxy core values) for a small sample in the very center of A2218.
Thus, it may be true that S0 bulges and ellipticals have a similar age and if S0's have a
prolonged episode of star formation, the evidence for this epoch will most likely be in their
disks and not their galaxy cores.  A difference of 0.2 mags in $uz-yz$ in Figure 8 would
correspond to color change from a 5 Gyrs population to a 13 Gyrs population (assuming that this
color represents the total luminosity of the galaxy of roughly solar metallicity), which
corresponds well with the Poggianti \etal estimate of 5 Gyrs minus 2 Gyrs of lookback time.

\subsection{Color Evolution of the Red Population}

The relatively low scatter in the $vz-yz$ CMR at high luminosities (Andreon 2003), and its
consistent slope from cluster to cluster (Bower \etal 1992, Aragon-Salamanca \etal 1993, Ellis
\etal 1997, Stanford, Eisenhardt \& Dickinson 1998, Kodama \etal 1998), presents an opportunity
to examine very small color differences in the red population due to lookback time, i.e. color
evolution.  Higher redshift clusters also offer the advantage of higher richnesses (meaning more
data) at the bright end of the CMR.  For example, the number of galaxies brighter than
$M_{5500}=-20$ in A2218 is about three times the number in Coma.  A2218 is at a redshift of
0.175 and A2125 is at a redshift of 0.247 which corresponds to a lookback times of 2.0 and 2.7
Gyrs ($H_o = 75$ km sec$^{-1}$ Mpc$^{-1}$, $\Omega_m=0.3$, $\Omega_\Lambda=0.7$).  The Bruzual
and Charlot (2003) spectrophotometric models, convolved to our multi-metallicity scheme,
predict a change in $vz-yz$ of 0.03 (blueward) and a increase in luminosity of 0.21 mags for a
time interval of 2.5 Gyrs (assuming that present day ellipticals in Coma are 13 Gyrs in mean
stellar age).

In previous papers, our analysis technique at this point was to ignore the changes in
luminosity and focus on the mean color of the red population.  This requires setting a limiting
magnitude to the sample, since fainter galaxies are bluer, and adopting a cutoff criteria to
divide the blue and red populations.  While the mean color is a robust measure of the red
population, the combined effects of color selected samples, bias at the limiting magnitude
and the CMR itself limit its usefulness.  With a sufficient number of galaxies, and a better
classification procedure, it should be possible to divide the red population into magnitude
bins and compare the CMR from averaged colors to the model predictions.

\begin{figure}
\centering
\includegraphics[width=16cm]{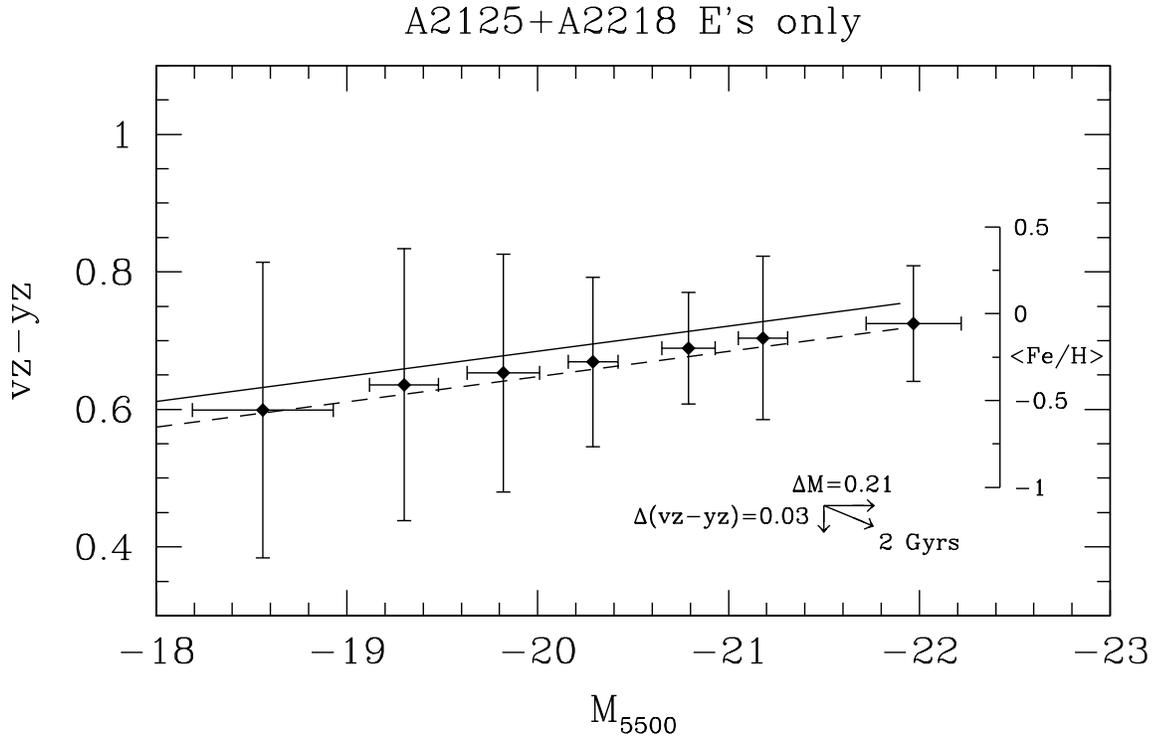}
\vspace{-5cm}
\caption{The metallicity CMR for class E galaxies in A2125 and A2218.  The data has been binned
into groups of 30 galaxies with the mean color and luminosity shown as solid symbols.  The
solid line is the CMR for Coma, the dashed line is the expected change in the CMR due to color
evolution for a 2 Gyrs lookback time.  Also shown is the metallicity scale from Rakos \&
Schombert (2004) for the $vz-yz$ color convolved to our multi-metallicity models.}
\end{figure}

Figure 9 presents just such an analysis for the red populations in A2125 and A2218 (combined).
Working in groups of 30 galaxies per bin produces the average $vz-yz$ colors found in Figure
9 where the error bars are the range of the data, not the error on the mean.  The solid line
is the CMR for the Coma/Fornax sample, the dotted line is the expected CMR for the changes in
luminosity and color as predicted by the SED models.  Also shown is an axis of mean Fe/H as
would be converted from $vz-yz$ using globular cluster calibrations.  

While the range per data point is large, all the binned data points lie below the Coma/Fornax
CMR in the expected direction for color evolution.  Note that each individual bin does not
carry sufficient statistical significance to support a claim of color evolution, but the entire
set of all luminosities is statistically significant and a new fit through the averaged data
produces a linear relation that is shifted almost exactly in the manner predicted by the
models for a passive evolving population.  While the large scale cosmological impact of this
result is small (the lookback time to A2125 and A2218 are too small to deduce the formation
epoch of the red population), the technique does provide some hope of using the CMR at very
high redshifts to study, in detail, the star formation history of ellipticals (Ellis \etal 1997).

\subsection{Blue Fraction}

The choice of A2125 and A2218 for deep narrow band imaging was made primarily because these two
rich clusters lie at similar intermediate redshifts, yet present contrasting global cluster
populations.  A2218 is denser than A2125 and has a more compact core population of red
galaxies, although the total richness for both clusters is similar.  A2125 is more irregular in
cluster structure (BM type II-III) than A2218 (BM type II), where cluster shape is considered
to be a measure of its dynamical state.  In the original work on the Butcher-Oemler effect,
A2125 has a higher blue fraction ($f_B$=0.17) compared to A2218 ($f_B$=0.11), although both
have blue fractions greater than the value for nearby clusters ($f_B$=0.04) and it is the blue
fraction that signals an evolutionary effect in intermediate redshift clusters.

Determining the blue fraction, $f_B$, is a selection driven process. The original $f_B$
criteria used the number of galaxies that were a set color difference from the ridgeline of the
cluster as a whole.  As discussed in Rakos \& Schombert (1995), the best color for selecting
the blue population in a cluster is $bz-yz$ since the CMR for $bz-yz$ is flat.  In any
metallicity sensitive color, such as $vz-yz$, a deeper limiting magnitude will introduce
relatively bluer (i.e. lower mean metallicity) galaxies.  For $bz-yz$, this effect is minimal
(although presumingly because there is a competing age effect for low mass ellipticals, see \S
3.5) and, through an analysis of low redshift spirals and irregulars, we have found that the
Butcher \& Oemler criteria corresponds to a color cutoff of $bz-yz$=0.2 in our rest frame
system (Rakos, Maindl \& Schombert 1996).  Applying this color criteria, and the using the same
limiting magnitude as Butcher \& Oemler ($M_{5500} < -20$), results in $f_B$ values of 0.15 and
0.06 for A2125 and A2218 respectfully.  This is similar to the original Butcher \& Oemler
values, with A2125 having a richer blue population than A2218.

\begin{figure}
\centering
\includegraphics[width=16cm]{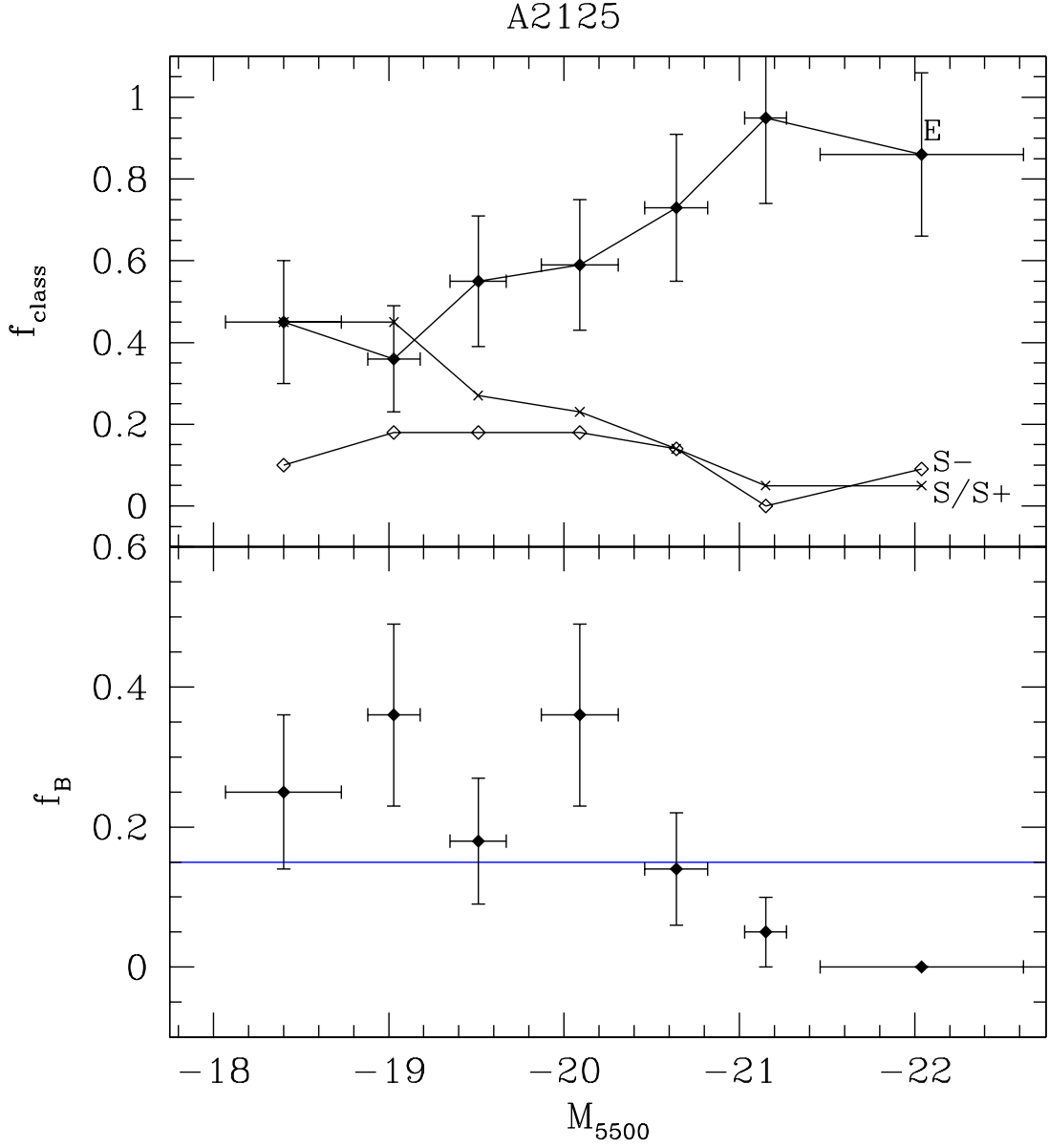}
\caption{The blue fraction ($f_B$) and photometric class fraction as a function of absolute
luminosity for A2125.  The blue line represents the cluster $f_B$ as defined by the
Butcher-Oemler criteria.  As noted in other Butcher-Oemler clusters, the blue population
increases its contribution at lower luminosities.  The top panel displays the run of
photometric classes with absolute magnitude.  A decrease in the E class is matched by an
increase in S and S+ class objects.  The lack of change in the S- population indicates that the
switch from star-forming to passive colors is abrupt.}
\end{figure}

\begin{figure}
\centering
\includegraphics[width=16cm]{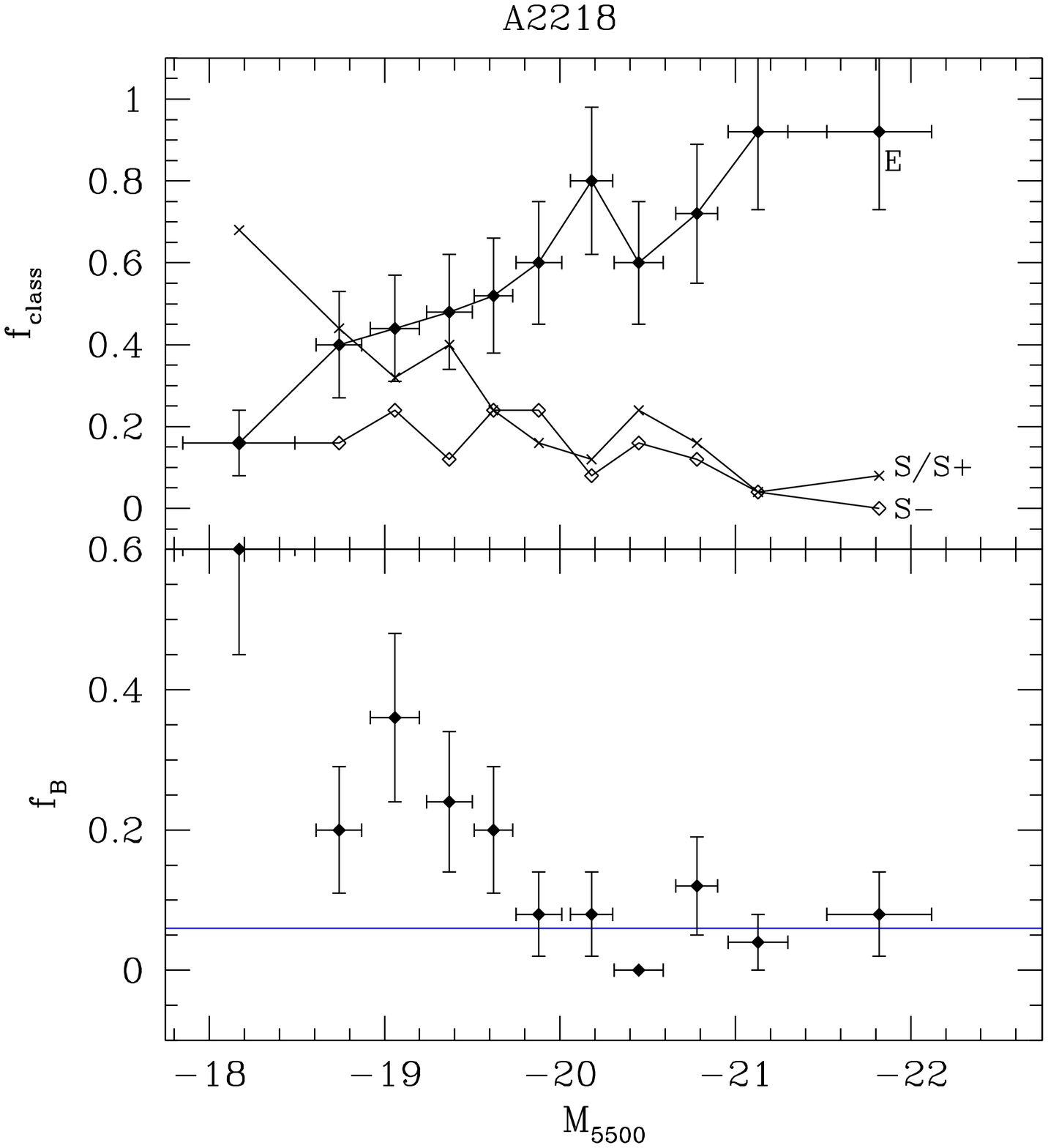}
\caption{The blue fraction ($f_B$) and photometric class fraction as a function of absolute
luminosity for A2218.  The blue line represents the cluster $f_B$ as defined by the
Butcher-Oemler criteria.  The trends are similar to Figure 10, with fewer blue galaxies near
the Butcher-Oemler cutoff of $M_{5500} = -20$.  This results in a lower global value of $f_B$
for A2218 compared to A2125 even though both clusters have a well developed population of low
luminosity blue galaxies.}
\end{figure}

A simple luminosity cutoff ignores the additional information offered by our four filter
photometry system.  Our data includes a much deeper sample of galaxies than the original
Butcher \& Oemler work and our filter system allows us finer discrimination by color.  To this
end, Figures 10 and 11 display the blue fraction, $f_B$, as a function of absolute luminosity
for A2125 and A2218 (bottom panels).  These values are based on the same color cutoff discussed
above, only applied to magnitude bins selected to have equal numbers of galaxies per bin.  The
blue line represents the total $f_B$ for the sample brighter than $M_{5500} = -20$, the
original Butcher-Oemler criteria.  The full sample displays some interesting trends.  For
example, both A2125 and A2218 follow the same pattern of an increasing blue fraction with
decreasing luminosity (Dahlen, Fransson \& Naslund 2001, De Propris \etal 2004).  In fact, when
using the entire sample, A2125 and A2218 have identical $f_B$ values of 0.19 due to the
increase in blue galaxies at low luminosities.  It is also obvious that the blue fraction for
the top three luminosity bins is very low, near 0.05, which is the value for Coma.  Therefore,
most of the blue population is composed of galaxies fainter than $-$21.5.

This result does not imply that cluster populations of A2125 and A2218 are identical to Coma at
the bright end of the luminosity function.  Even though the blue fraction is low above $-$21.5,
there are a significant number of bright blue galaxies (see CMR in Figures 6 and 7) that have no
counterparts in Coma (see Figure 3 of Odell, Schombert \& Rakos 2002).  For comparison, the
same analysis can be performed using photometric classifications as discussed in \S 3.1.  The top
panels in Figures 10 and 11 display the fraction of E, S-, S and S+ type galaxies as a function
of absolute luminosity.  The trend here is clearer than with the blue fraction such that there is a
steady decrease in the fraction of E types with decreasing luminosity matched by an increase in
S and S+ types.  The transition S- type remains nearly constant.  This trend is identical in
A2125 and A2218, despite the differences in the density and richness.  

Our interpretation of these trends in color and photometric class derives from comparison with
nearby clusters (Coma and Fornax) and several intermediate redshift clusters with strong blue
populations (A115, A2283, A2317, Rakos \etal 2000).  In the intermediate redshift clusters, the
blue population divides itself into two sub-populations, 1) a bright population with spiral
colors and 2) a fainter, dwarf starburst population (SFR $>$ 10 $M_{\sun}$ yr$^{-1}$).
Consistently, for all the intermediate redshift clusters we have studied, there is a deficiency
in blue galaxies in the luminosity range from $-$20 to $-$21 compared to the red population,
hence a decrease in $f_B$ over these luminosities.  Both A2125 and A2218 differ from this trend
by lacking a dominate, bright blue population ($M_{5500} < -20$), although they still have a
number of starburst galaxies (e.g. C153).  In the next section, we will explore the properties
of the blue population in order to understand its origin.

\subsection{Properties of the Blue Populations}

Given the similarity in trends found in Figures 10 and 11, the question arises of how do the two
clusters distinguish themselves with respect to their blue fractions and the state of the
Butcher-Oemler effect.  Most photometric surveys do not sample deep enough into the luminosity
function to allow the faint blue population to significantly contribute to the $f_B$ value,
although certainly some of the range and scatter in $f_B$ from cluster to cluster is due to
differing magnitude cutoffs and different filter systems.  Inspection of the CMR for A2125 and
A2218 in Figures 6 and 7 show that the blue galaxies at the high end of the luminosity function
are due to galaxies with spiral-like colors in A2218, however, in A2125, we find a majority of
the brightest blue galaxies have starburst colors.

This can also been seen in the physical morphology of the blue population from the WFPC2
images.  In Table 2, it can be seen that all the blue galaxies in A2125 are late-type Sc and
Irr's.  However, in A2218, a majority of the blue galaxies (60\%) are early-type S0's or Sa/Sb
by morphology.  Figures 12 and 13 presents a mosaic of a few of the brightest blue galaxies in
A2125 and A2218.  The blue galaxies in A2218 follow the trends for the Hubble sequence in
nearby galaxies, meaning the bluest systems have the latest Hubble classes.  For example, \#22
and \#481 are early-type spirals, \#75 and \#164 are late-type.  The blue galaxies in A2125, in
general, are more disturbed from the standard Hubble sequence.  Galaxy \#481 in A2125 is the
well studied starburst C153 (Wange \etal 2004) whose optical and HI morphology is distorted as
it plows into the cluster ICM.  The primary difference between A2125 and A2218 versus Coma is
that fact that there are no counterparts in Coma to galaxies shown in Figures 12 and 13.  The
closest Coma galaxy with colors of a star-forming object has a luminosity of only $-$18.5
(galaxy A014 in Odell, Schombert \& Rakos 2002).

\begin{figure}
\centering
\includegraphics[width=16cm]{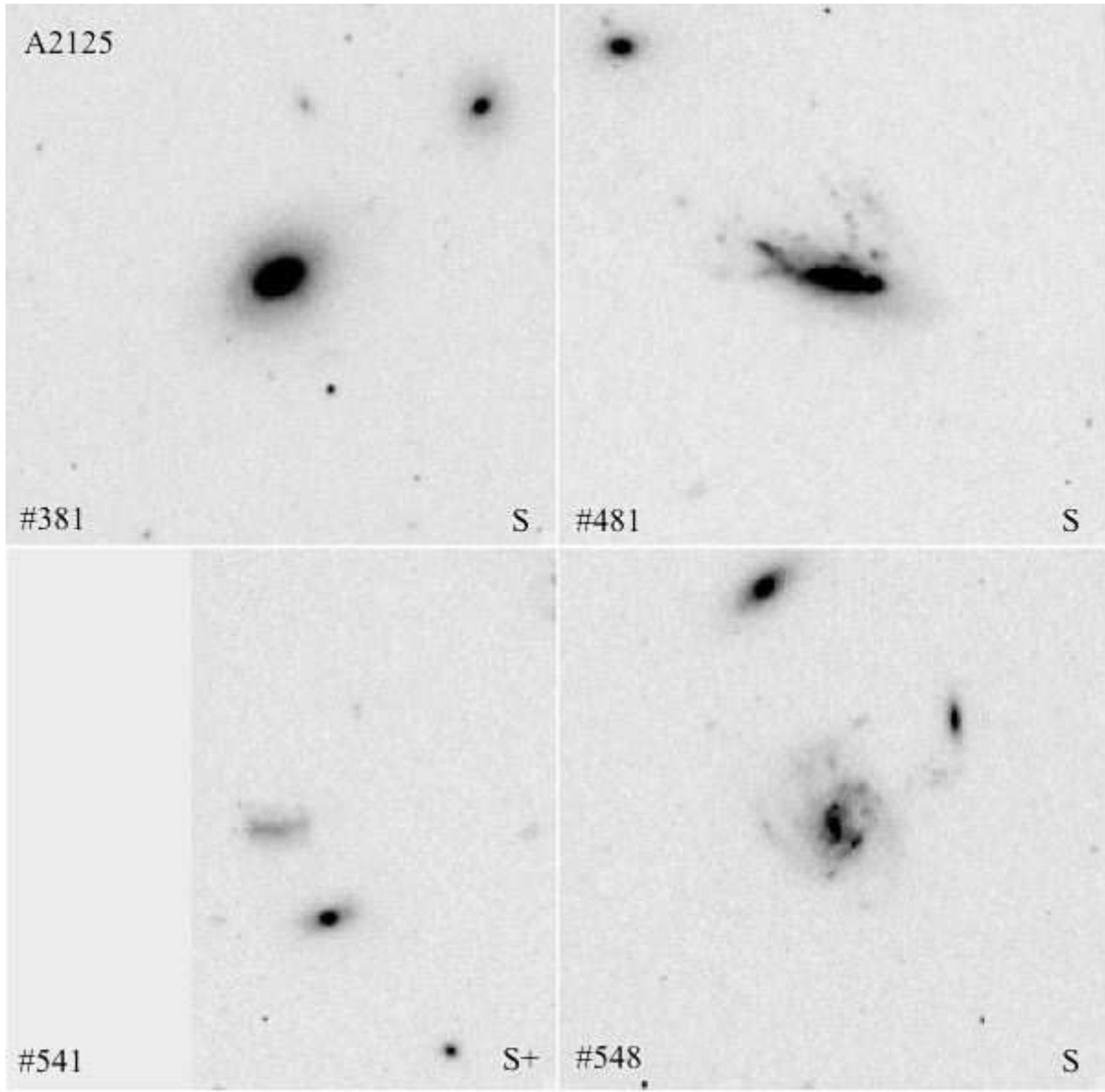}
\caption{A mosaic of four of the brighter blue galaxies in A2125 from WFPC2 imaging (F606W).
Each subimage is 50 kpc wide.  Galaxy's \#381 and \#481 are early-type spirals, \#541 and \#548
are late-type.  The blue galaxies in A2125, in general, are more disturbed from the standard
Hubble sequence.  Galaxy \#481 in A2125 is the well studied starburst C153 (Wange \etal 2004)
whose optical and HI morphology is distorted as it plows into the cluster ICM.}
\end{figure}

\begin{figure}
\centering
\includegraphics[width=16cm]{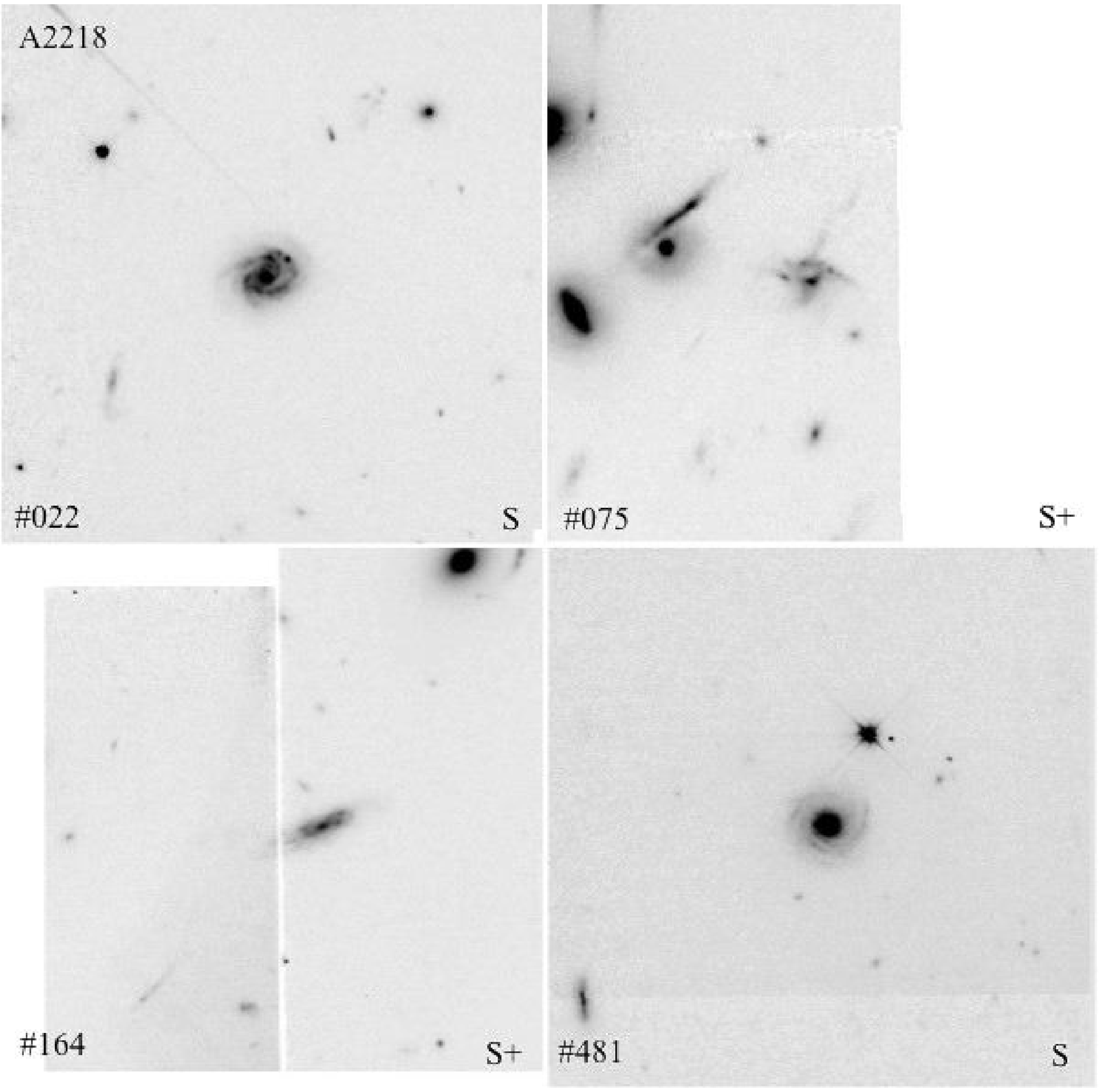}
\caption{A mosaic of four of the brighter blue galaxies in A2218 from WFPC2 imaging (F606W).
Each subimage is 50 kpc wide.  The blue galaxies in A2218, in general, are follow the standard
Hubble sequence, galaxy \#075 is an expection to this rule being a highly disturbed,
interacting system.}
\end{figure}

This is the essence of the Butcher-Oemler effect, meaning the difference between clusters with
an active Butcher-Oemler population (i.e. A2125) and ones which display a higher fraction of
galaxies with spiral-like star formation rates simply due to cosmological time (i.e. A2218).
For A2125, there are many indications that the Butcher-Oemler population is
linked to the dynamical state of the cluster by the interaction between gas-rich galaxies and
the cluster environment either through gas stripping (Gunn \& Gott 1972) or the cluster tidal
field (Moore \etal 1996).  Compared to A2218, a more relaxed cluster, the difference in their
blue populations lies in the late-type, starburst systems, such as galaxy \#481, which have
strong evidence of environmentally induced star formation.  Removing the starburst systems from
A2125 would produce blue fractions similar to A2218.  This is not to say that the blue
population in A2218 is insignificant for, as with A2125, the blue galaxies in Figure 12 have no
counterparts in the core of Coma.  However, the blue galaxies in their appearance are
consistent with present-day Hubble types and their colors match normal spirals.  So even at
lookback times of only 2 Gyrs ago, there are significant differences in the cluster populations
of rich clusters both in the properties of the S0's (higher fraction of bulge+disk systems) and
the blue galaxies (higher fraction of late-type systems).

\begin{figure}
\centering
\includegraphics[width=16cm]{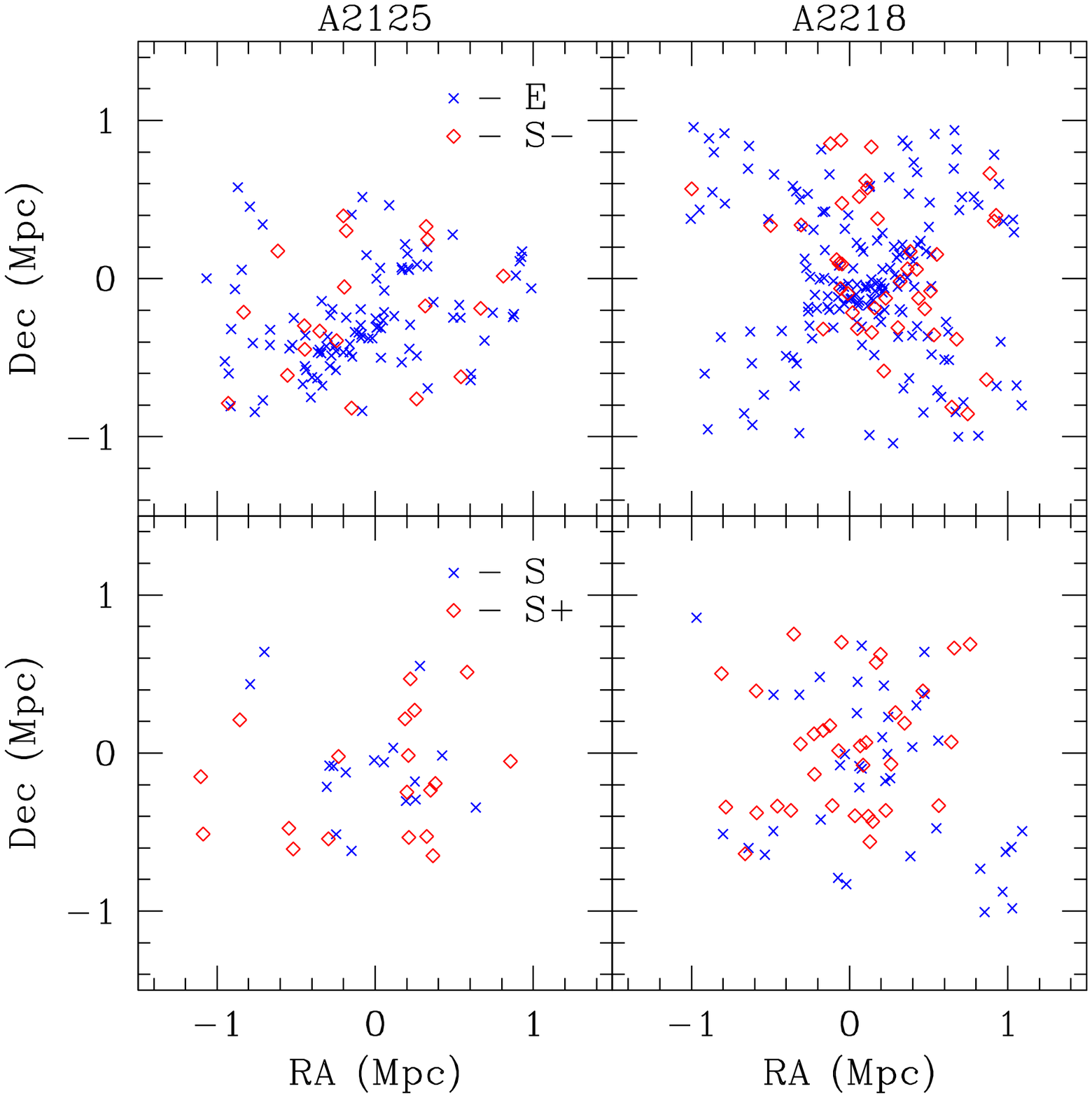}
\caption{The position location of all four photometric classes in A2125 and A2218.  Only the E
class displays a clear central concentration, the other three classes are evenly distributed
throughout the clusters.}
\end{figure}

The high richness of A2125 and A2218 also allows us to test the change in galaxy population
with cluster radial distance.  Figure 14 displays the distribution of different galaxies types
as projected onto the sky.  In both clusters it is obvious that the early-type galaxies (E and
S-) are strongly clustered in the core and that the late-type galaxies (S and S+) are more
evenly distributed.  This has been demonstrated in numerous papers (Abraham \etal 1996,
Ellingson \etal 2001) that the blue population is more dense in the outlying regions of a
cluster, presumably infalling field galaxies.  Thus, one origin hypothesis to the blue
population is that they are composed of gas-rich field systems which infall into a dense
cluster and environmental influences produce a starburst. However, notable counterexamples,
such as galaxy \#481 (aka C153 in A2125), indicate that some of the starburst systems near the
cluster core are undergoing of a strong interaction with the dense ICM and are not simply an
infalling field population.

\begin{figure}
\centering
\includegraphics[width=16cm]{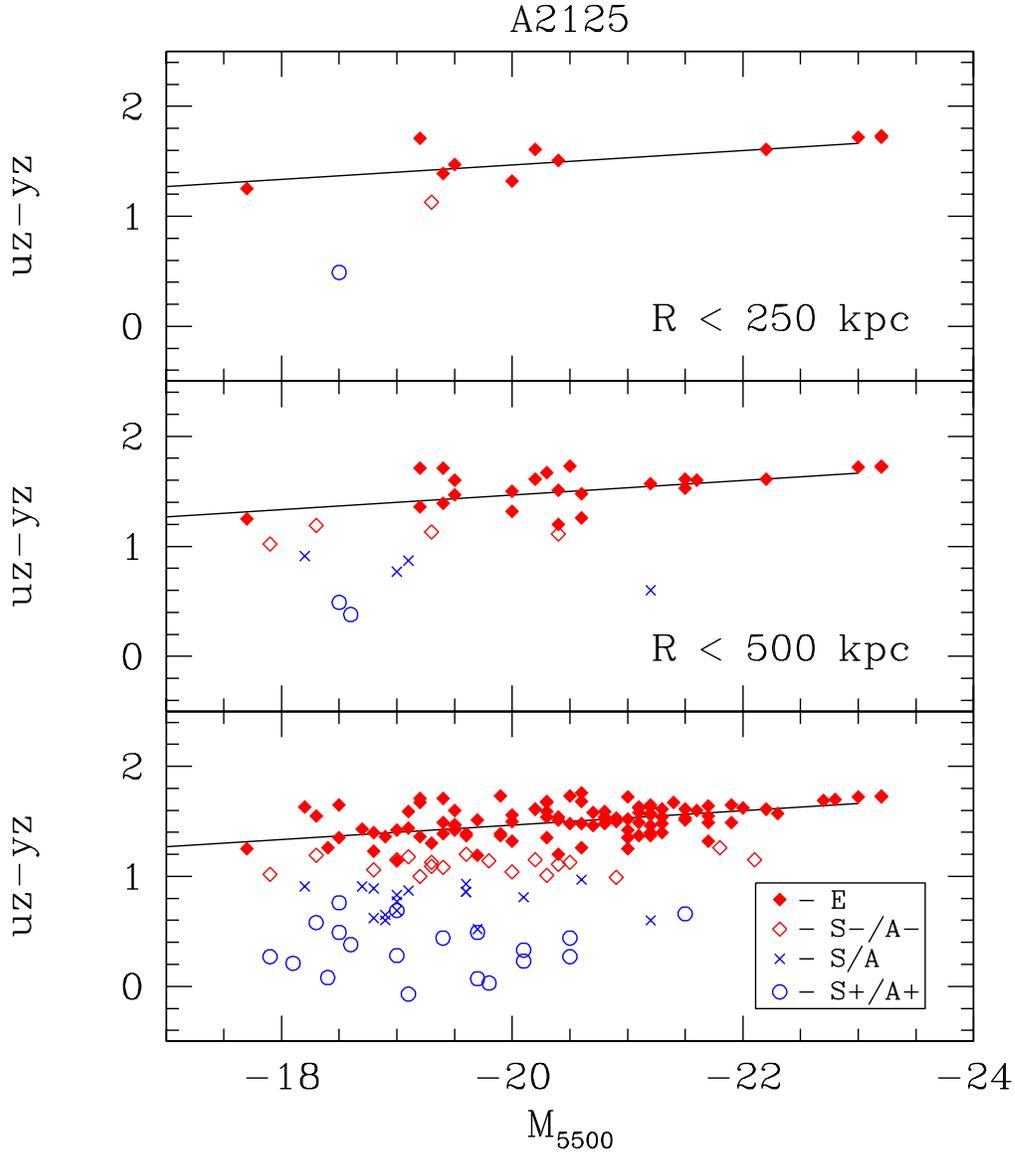}
\caption{The $uz-yz$ CMR as a function of cluster distance in A2125.  The increase in the
number of blue galaxies with radius is evdent.  There is no change in the red population with
radius with respect to the CMR slope, although an increase number of tranisition objects (S-)
would complicate the fit by morphology alone.}
\end{figure}

\begin{figure}
\centering
\includegraphics[width=16cm]{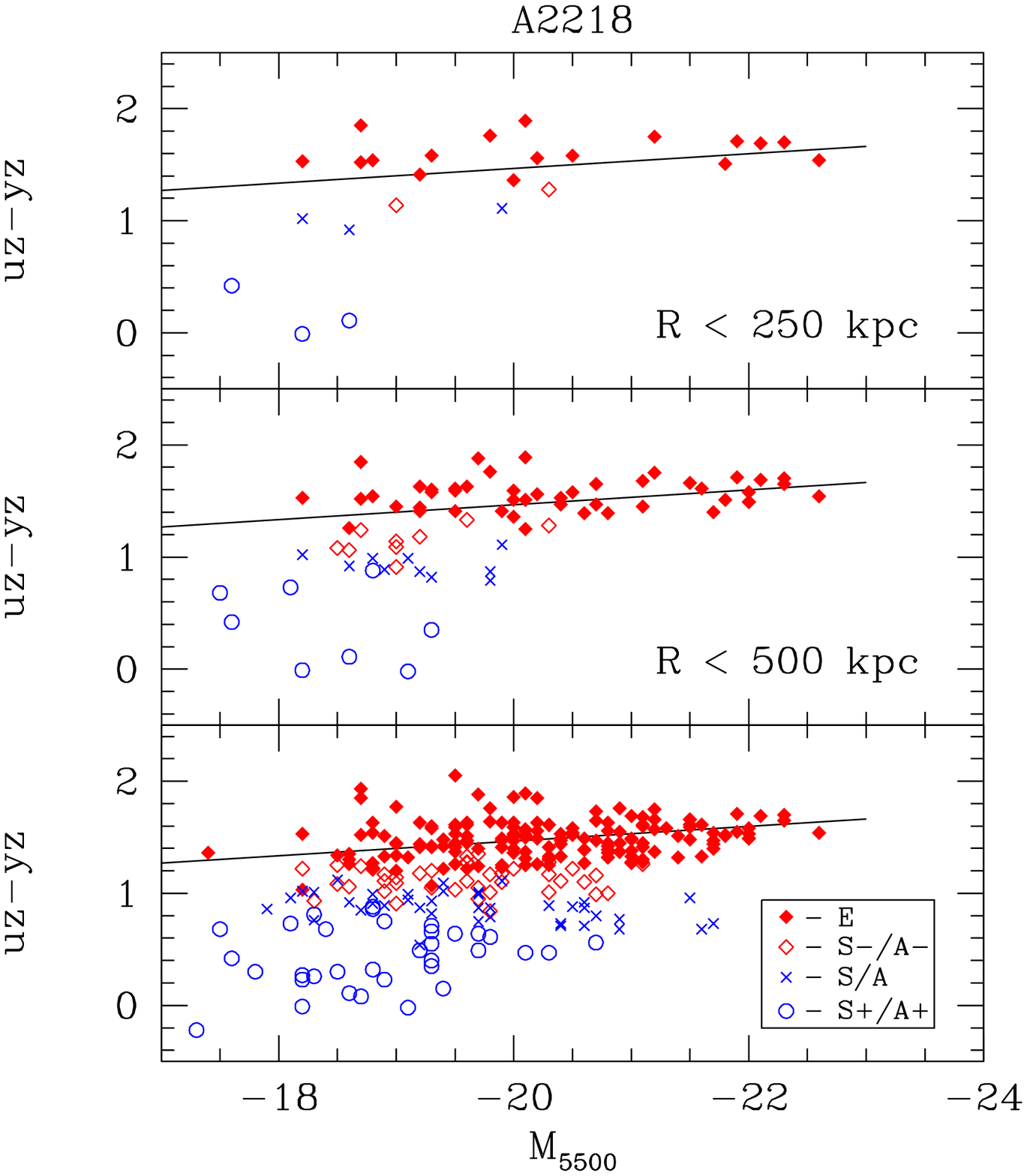}
\caption{The $uz-yz$ CMR as a function of cluster distance in A2218.}
\end{figure}

The radial distance effect on the blue population can be seen in more detail in Figures 15 and
16, the $uz-yz$ CMR divided into three bins of radial distance from the center center.  Both
A2125 and A2218 display a lack of any bright blue galaxies in their cluster cores, with an
increase in the number of blue galaxies, and their mean luminosity, at larger radii.  We note
that these trends are identical in both clusters despite their differences in cluster structure
and dynamical state.  In fact, aside from a slightly higher fraction of starburst systems in
A2125, neither cluster can be distinguished from each other based on the distribution of their
blue populations.

\section{CONCLUSIONS}

One of the main obstacles in the study of galaxy evolution is the wide range of environmental
effects on the star formation and morphological history of galaxies.  Given the diversity of
environments in which galaxies are found, and the importance that environment plays in our
galaxy formation scenarios, predicting the evolution of particular systems may elude our
understanding for the near future.  In this study, we have attempted to minimize the range of
evolutionary effects by comparing two rich clusters at similar redshifts, yet relatively short
lookback times, to dense present-day clusters (i.e. Coma).  While investigating galaxy
populations at redshifts of 0.2 will not reveal a great deal of information concerning
formation epoch, it does allow us to test passive evolution models and determine the
characteristics of this epoch's dying blue population.

We divide our results into three parts: 1) the raw observables, 2) model dependent
interpretation of those observables and 3) speculation on the meaning of the data.  The first
two, our primary observables and interpretation of the data, are summarized in the following
list:

\begin{itemize}

\item{} Our four color filter system allows the spectrophotometric classification of the galaxy
sample using principal component analysis.  We have broken the sample into four color groups,
E, S-, S and S+ in order of increasing blueness due to star formation.  A non-thermal color
contribution can also be detected in the PC2 axis (class A), although, not surprisingly, these
objects are very rare in our sample.  The classification results are summarized in Table 1,
A2125 and A2218 are very similar in their photometric properties, both being richer in blue
galaxies compared to Coma.  Even the core regions of A2125 and A2218 have a higher fraction of
blue galaxies compare to the core of Coma, despite the higher galaxy densities for A2125 and
A2218.

\item{} Archive WFPC2 imaging allows the morphological classification of a subset of the
photometric sample (32\% of the total sample).  Typically, these galaxies are in the cluster
cores due to observing constraints.  The relationship between visual morphology and photometric
classification is shown in Table 2.  As expected, there is a tight correlation between
photometric classification by color and morphology.  Only a few early-type disk galaxies are
classified as red (E or S-) and all the blue systems display late-type morphology.  It is not
the case that the blue galaxies are fading into large bulge S0's, as most are late-type systems
(see also Couch \etal 1998).  There is a noticeable deficiency in ellipticals (r$^{1/4}$ shaped
objects) of about 15\% compared to Coma, which matches in increased number of late-type
galaxies.  However, the differences in the S0 population (see below) indicate that the
conversion of S0's to ellipticals is matched by the conversion of late-type galaxies to S0's.

\item{} Detailed surface photometry could be performed on the WFPC2 images down to 28 mag
arcsecs$^{-2}$ and the resulting surface brightness profiles are divided into three types; 1)
r$^{1/4}$ shape profiles (B), 2) combination of r$^{1/4}$ shape bulge and exponential disk
(B+D) and 3) pure exponential disk (D).  Table 3 and 4 summarizes the relationship between
surface brightness profile classification and photometric classification and morphology.
As expected, the blue population is almost completely composed of disk galaxies with a few
bulge+disk objects.  The red population is composed of a mixture of B, B+D and D systems in a
average ratio of 25\%:40\%:10\% of the total cluster population.  This differs significantly
from Coma's 40\%:20\%:30\% mixture.  An increase of red pure disk galaxies (lenticulars) may
arise from faded blue galaxies; however, the increase in B systems and the decrease in B+D
systems can only arise if 1/2 of the S0's at intermediate redshift are converted, structurally,
into ellipticals.  This can also be seen in Table 4 where approximately 80\% of the S0's in
A2125 and A2218 are bulge+disk systems, whereas only 35\% of the S0's in Coma have this type of
structure.  The majority of S0's in nearby clusters are lenticulars (van den Bergh 1994), 
which are a minority at intermediate redshifts.

\item{} The color distributions for the cluster galaxies occupies the same phase space as SDSS
galaxies (Smolcic \etal 2004) with barely detectable evolutionary effects.  The red population
follows a strict metallicity-color relation (as compared to SED models).  The blue population
has colors consistent with nearby galaxies undergoing star formation at rates varying from 1 to
10 $M_{\sun}$ per yr.  None of the blue galaxies are composed of a pure young stellar population,
but rather follow the `frosting' models, a small, young population on top of an older stellar
population (i.e. recent star formation).

\item{} The color-magnitude relation for red galaxies follows the same slope and zeropoint as
Coma with minor, but detectable, deviations that correspond to direct evidence for color
evolution.  This consistence to the CMR, plus low scatter is also seen in clusters up redshifts
of 0.5 (Ellis \etal 1997), places fairly strong constraints on the epoch of star formation
for the red population to be greater than a redshift of 5 (the red envelope, see Rakos \&
Schombert 1995).  The lack of a meaningful slope to the $bz-yz$ color argues for an age effect
in the direction of older mean age, or later epoch of formation, for lower luminosity
ellipticals (see Rakos \& Schombert 2004).

\item{} The calculated blue fraction ($f_B$) varys widely depending on the luminosity and
radius cutoffs.  Using the original Butcher-Oemler criteria derives a blue fraction of 0.15 for
A2125 and 0.06 for A2218, both are higher than typical values for present-day rich clusters.
The blue fraction increases with decreasing luminosity and with increasing radius.  Both
clusters have similar $f_B$ values for the high end of the luminosity function; however, the
type of blue galaxies differs between the clusters.  In A2218, the colors of these bright blue
galaxies indicate normal, spiral-like star formation rates and their morphology from WFPC2
images is consistent with early-type disks.  The opposite is true in A2125, most of the
brightest blue galaxies display starburst colors and later morphological types.  Thus, the
Butcher-Oemler effect is one of both color and morphological evolution.

\item{} An increasing
number of starburst colors are found at lower luminosities in both clusters indicating that the
Butcher-Oemler population is divided into two sub-populations, a bright, fading spiral
population and a faint, starburst dwarf population.  We note that interpretations of star
formation rate are based on continuum colors (i.e. averaged over the past couple of Gyrs),
spectroscopic information at this redshift indicates that spirals are `quenched' in that their
emission lines such that current SFR values are lower than normal spirals (Couch \etal 1998,
Dressler \etal 1999, Poggianti \etal 1999).  It is this comparison between spectral values
and continuum colors that makes photometry of distant clusters a necessary component to
understanding the evolution of the blue population.

\end{itemize}

In many ways, the galaxies in A2125 and A2218 are similar to those found in present-day rich
clusters.  The cores of both clusters are dominated by red, elliptical galaxies despite the
clear dynamical differences between the irregular A2125 and the cD dominated A2218.  The red
populations (ellipticals and S0's) follow the same mass-metallicity relationship as Coma and
Fornax and even display the correct amount of evolution in color and luminosity expected from
passive evolution models.  It is interesting to note that both Coma and Fornax, in the present
epoch, and A2125 and A2218, two Gyrs in the past, have a flat $bz-yz$ CMR.  The SED models
predict a small but detectable slope from the metallicities implied by the $uz-yz$ and $vz-yz$
colors.  The lack of a slope in Figures 6 and 7 imply an age effect that is coupled to the
metallicity of a galaxy (the direction is such that metal-poor, or low mass, galaxies are
older).  If an age effect is responsible for the $bz-yz$ CMR, then the estimated age difference
from low to high mass is small, less than 2 to 3 Gyrs (see Rakos \& Schombert 2004).

A change in mean age of a stellar population with mass has three possible interpretations: 1)
lower mass galaxies form before high mass galaxies, 2) all galaxies form at the same time, but
high mass galaxies have an extended period of star formation that produces a younger mean age
or 3) higher mass galaxies acquire a younger stellar population through mergers or cannibalism.  
Hierarchical models of galaxy formation favor the first scenario; however, the mass-metallicity
relation, seen so clearly in the $uz-yz$ and $vz-yz$ CMR, makes a process of building high
metallicity, high mass galaxies from low metallicity, low mass galaxies problematic.  Certainly
more gas mass in high mass galaxies implies a longer star formation epoch, but the CMR is best
explained by an abrupt halt to star formation by galactic winds and, thus, extending the time of star
formation beyond a single Gyr is also difficult to match to the CMR.  Lastly, galaxy
cannibalism is present in clusters (Schombert 1987), but the number of star-forming galaxies in a
cluster core to accrete is limited and previous work (Kuntschner 2000, Trager \etal 2000)
has demonstrated that cluster ellipticals are as old as globular clusters with very few examples of
ellipticals with a younger population mixed into the original stellar population.

In agreement with the conclusions on distant clusters by Dressler \etal (1997), the red
population in A2125 and A2218 appears to predate the formation of the cluster environment and
its color properties appear to be independent of the cluster dynamical state.  The key difference
between the cluster populations in A2125 and A2218 lies in the properties of the blue
population.  Although this was only a sample of two clusters, the more dynamically relaxed (A2218)
cluster's blue population has fewer starburst systems than the less dynamically evolved (A2125)
cluster.  And many of those starburst galaxies show clear signs of environmentally
induced star formation (see Wange \etal 2004) such that the age of the stellar population in
blue cluster galaxies is linked to the dynamical state of the cluster.  

When we compare all the galaxy properties (color and structure) we also find differences in the
S0 population.  A majority of the S0's in A2125 and A2218 are B+D type galaxies by structure,
meaning they have detectable bulge and disk combinations.  However, these types of S0's are
rare in Coma where a majority of the S0 population is comprised of pure disk systems (i.e.
lenticulars).  A clue to where the B+D S0's have gone lies in Figure 2, a plot of the structure
parameters of the r$^{1/4}$ parts of the red population (ellipticals and bulges).
The structure of elliptical galaxies in A2125 and A2218 follow the same effective radius
($r_e$) versus effective surface brightness ($\mu_e$) relations as nearby cluster ellipticals.
The bulges of B+D S0's also follow the elliptical relation (i.e. they are structurally
identical to ellipticals).   This suggests that many of the S0's at intermediate redshift are
stripped of their disks (either by encounters with other galaxies or by the cluster tidal
field) and their remaining bulges become ellipticals.  

Lastly, we comment on the apparent special time that we live in with respect to galaxy
evolution.  One of the philosophical difficulties with the Butcher-Oemler effect is that we
appear to live at just the right epoch were star formation has (recently) ended in cluster
cores.  This would appear, at first, to be a violation of the Copernican principle, applied in
a temporal fashion.  However, the Universe does evolve.  Certainly, we expect to find observers
only at a cosmic epoch that can support complex lifeforms (Davies 2004).  Once a galaxy's gas
supply is exhausted, it will rapidly take on a smooth appearance and red colors (i.e. S0-like).
Thus, the Butcher-Oemler effect is merely a statement of the gas supply in various clusters and
the rapid nature that various cluster environmental effects can have on a particular galaxy.
It is also a mistake to assume that the Butcher-Oemler effect is one of decreasing blue
fraction to present-day values.  There are numerous present-day clusters with large blue galaxy
populations (e.g. Hercules A2151) and the evolutionary process is not finished in those
clusters.  There is insufficient information on the projected course of cluster populations to
conclude that today represents a final state in cluster evolution.

\acknowledgements

Financial support from Austrian Fonds zur Foerderung der Wissenschaftlichen Forschung and NSF
grant AST-0307508 is gratefully acknowledged.  Some of the data presented in this paper were
obtained from the Multimission Archive at the Space Telescope Science Institute (MAST). STScI
is operated by the Association of Universities for Research in Astronomy, Inc., under NASA
contract NAS5-26555. Support for MAST for non-HST data is provided by the NASA Office of Space
Science via grant NAG5-7584 and by other grants and contracts.  This research has made use of
the NASA/IPAC Extragalactic Database (NED) which is operated by the Jet Propulsion Laboratory,
California Institute of Technology, under contract with the National Aeronautics and Space
Administration.  We thank KPNO for the usual dribble of telescope time for this project and the
mountain staff for their efforts during the observing runs.  We also thank the anonymous
referee for his/her time in proofing this paper.

\clearpage
\pagestyle{empty} 
\renewcommand{\arraystretch}{.6}
\begin{deluxetable}{lccccc}
\tablecolumns{6}
\small
\tablewidth{0pt}
\tablecaption{Population Fractions by Photometric Classification}
\tablehead{
\colhead{Type} &
\colhead{A2125} &
\colhead{A2125 (core)$^a$} &
\colhead{A2218} &
\colhead{A2218 (core)$^a$} &
\colhead{Coma (core)$^a$} \nl
}
\startdata
E     & 0.64 & 0.73 & 0.60 & 0.72 & 0.93 \nl
S-/A- & 0.12 & 0.11 & 0.14 & 0.10 & 0.02 \nl
S/A   & 0.11 & 0.11 & 0.14 & 0.10 & 0.04 \nl
S+/A+ & 0.13 & 0.05 & 0.12 & 0.08 & 0.02 \nl
\enddata
\tablecomments{
$^a$ R $<$ 500 kpc
}
\end{deluxetable}

\begin{deluxetable}{lrrrrr}
\tablecolumns{6}
\small
\tablewidth{0pt}
\tablecaption{WFPC2 Population Fractions by Morphology and Photometric Classification}
\tablehead{
\colhead{Type} &
\colhead{E} &
\colhead{S0} &
\colhead{Sa/Sb} &
\colhead{Sc/Irr} &
\colhead{Totals} \nl
\colhead{} &
\colhead{(\%)} &
\colhead{(\%)} &
\colhead{(\%)} &
\colhead{(\%)} &
\colhead{(\%)} \nl
}
\startdata
\multicolumn{6}{c}{\underline{A2125}}\nl
\tablevspace{0.2cm}
E      & 23.1 &57.7 & 7.7 & 0.0 &88.5 \nl
S-/A-  &  0.0 & 0.0 & 0.0 & 0.0 & 0.0 \nl
S/A    &  0.0 & 0.0 & 0.0 & 7.7 & 7.7 \nl
S+/A+  &  0.0 & 0.0 & 0.0 & 3.8 & 3.8 \nl
       & 23.1 &57.7 & 7.7 &11.5 & \nl
&&&&&\nl
E/S-/A- &23.1 &57.7  &7.7 & 0.0 &88.5\nl
S/S+    & 0.0 & 0.0  &0.0 &11.5 &11.5\nl
A/A+    & 0.0 & 0.0 & 0.0 & 0.0 & 0.0\nl
&&&&&\nl
\multicolumn{6}{c}{\underline{A2218}}\nl
\tablevspace{0.2cm}
E      & 25.2 &37.4 & 0.0 & 0.0 &62.6 \nl
S-/A-  &  1.9 &13.1 & 0.9 & 0.0 &15.9 \nl
S/A    &  0.0 & 6.5 & 5.7 & 0.9 &13.1 \nl
S+/A+  &  0.9 & 0.0 & 0.0 & 7.5 & 8.4 \nl
       & 28.0 &57.0 & 6.6 & 8.4 & \nl
&&&&&\nl
E/S-/A- &27.1 &50.5 & 0.9 & 0.0 &78.5\nl
S/S+    & 0.0 & 2.8 & 4.7 & 4.7 &12.2\nl
A/A+    & 0.9 & 3.7 & 1.0 & 3.7 & 9.3\nl
&&&&&\nl
\multicolumn{6}{c}{\underline{Coma}}\nl
\tablevspace{0.2cm}
E      & 40.0 &48.3 & 1.7 & 0.0 &90.0 \nl
S-/A-  &  0.0 & 1.7 & 0.0 & 0.0 & 1.7 \nl
S/A    &  3.3 & 1.7 & 1.7 & 0.0 & 6.7 \nl
S+/A+  &  0.0 & 1.6 & 0.0 & 0.0 & 1.6 \nl
       & 43.3 &53.3 & 3.4 & 0.0 & \nl
&&&&&\nl
E/S-/A- &40.0 &50.0 & 1.7 & 0.0 &91.7\nl
S/S+    & 1.6 & 0.0&  1.7 & 0.0 & 3.3\nl
A/A+    & 1.7 & 3.3&  0.0 & 0.0 & 5.0\nl
\enddata
\end{deluxetable}

\begin{deluxetable}{lrrrr}
\tablecolumns{5}
\small
\tablewidth{0pt}
\tablecaption{WFPC2 Population Fractions by Surface Brightness Profile and Photometric Classifications}
\tablehead{
\colhead{Type} &
\colhead{B} &
\colhead{B+D} &
\colhead{D} &
\colhead{Totals} \nl
\colhead{} &
\colhead{(\%)} &
\colhead{(\%)} &
\colhead{(\%)} &
\colhead{(\%)} \nl
}
\startdata
\multicolumn{5}{c}{\underline{A2125}}\nl
\tablevspace{0.2cm}
E      & 23.1 &53.8 &11.6 &88.5 \nl
S-/A-  &  0.0 & 0.0 & 0.0 & 0.0 \nl
S/A    &  0.0 & 0.0 & 7.7 & 7.7 \nl
S+/A+  &  0.0 & 0.0 & 3.8 & 3.8 \nl
       & 23.1 &53.8 &23.1 & \nl
&&&&\nl
E/S-/A- &23.1 &53.8 &11.5 &88.5 \nl
S/S+    & 0.0 & 0.0 &11.5 &11.5 \nl
A/A+    & 0.0 & 0.0 & 0.0 & 0.0 \nl
&&&&\nl
\multicolumn{5}{c}{\underline{A2218}}\nl
\tablevspace{0.2cm}
E      & 25.2 &34.6 & 2.8 &62.6\nl
S-/A-  &  1.9 & 9.3 & 4.7 &15.9\nl
S/A    &  0.0 & 7.5 & 5.6 &13.1\nl
S+/A+  &  0.9 & 0.9 & 6.5 & 8.4\nl
       & 28.0 &52.3 &19.6 &\nl
&&&&\nl
E/S-/A- &27.1 &43.9 & 7.5 &78.5\nl
S/S+    & 0.0 & 3.7 & 8.4 &12.1\nl
A/A+    & 0.9 & 4.7 & 3.7 & 9.3\nl
&&&&\nl
\multicolumn{5}{c}{\underline{Coma}}\nl
\tablevspace{0.2cm}
E      & 40.0 &18.3 &31.7 &90.0 \nl
S-/A-  &  0.0 & 0.0 & 1.7 & 1.7 \nl
S/A    &  3.3 & 1.7 & 1.7 & 6.7 \nl
S+/A+  &  0.0 & 0.0 & 1.7 & 1.7 \nl
       & 43.3 &20.0 &36.7 & \nl
&&&&\nl
E/S-/A- &40.0 &18.3 &33.3 &91.7 \nl
S/S+    & 1.7 & 0.0 & 1.7 & 3.3 \nl
A/A+    & 1.7 & 1.7 & 1.7 & 5.0 \nl
\enddata
\end{deluxetable}

\begin{deluxetable}{lrrrr}
\tablecolumns{5}
\small
\tablewidth{0pt}
\tablecaption{WFPC2 Population Fractions by Surface Brightness Profile and Morphology}
\tablehead{
\colhead{Type} &
\colhead{B} &
\colhead{B+D} &
\colhead{D} &
\colhead{Totals} \nl
\colhead{} &
\colhead{(\%)} &
\colhead{(\%)} &
\colhead{(\%)} &
\colhead{(\%)} \nl
}
\startdata
\multicolumn{5}{c}{\underline{A2125}}\nl
\tablevspace{0.2cm}
E      & 23.1 & 0.0 & 0.0 &23.1 \nl
S0     &  0.0 &50.0 & 7.7 &57.7 \nl
Sa     &  0.0 & 3.8 & 3.9 & 7.7 \nl
Sc     &  0.0 & 0.0 &11.5 &11.5 \nl
       & 23.1 &53.8 &23.1 & \nl
&&&&\nl
\multicolumn{5}{c}{\underline{A2218}}\nl
\tablevspace{0.2cm}
E     &  28.0 & 0.0 & 0.0 &28.0 \nl
S0    &   0.0 &46.7 &10.3 &57.0 \nl
Sa    &   0.0 & 4.7 & 1.8 & 6.5 \nl
Sc    &   0.0 & 0.9 & 7.5 & 8.4 \nl
      &  28.0 &52.3 &19.6 & \nl
&&&&\nl
\multicolumn{5}{c}{\underline{Coma}}\nl
\tablevspace{0.2cm}
E      & 43.3 & 0.0 & 0.0 &43.3 \nl
S0     &  0.0 &18.3 &35.0 &53.3 \nl
Sa     &  0.0 & 1.7 & 1.7 & 3.4 \nl
Sc     &  0.0 & 0.0 & 0.0 & 0.0 \nl
       & 43.3 &20.0 &36.7 & \nl
\enddata
\end{deluxetable}


\begin{references}

\reference{} Abraham, R., va, den, Glazebrook, K., Ellis, R., Santiago, B., Surma, P. \& Griffiths, R. 1996, \apjs, 107, 1
\reference{} Abraham, R. 1999, \apss, 269, 323
\reference{} Andreon, S. 1998, \apj, 501, 533
\reference{} Andreon, S. 2003, \aap, 409, 37 
\reference{} Andreon, S. 1998, \apj, 501, 533
\reference{} Aragon-Salamanca, A., Ellis, R., Couch, W. \& Carter, D. 1993, \mnras, 262, 764
\reference{} Arimoto, N. \& Yoshii, Y. 1987, \aap, 173, 23
\reference{} Bender, R., Burstein, D. \& Faber, S. 1992, \apj, 399, 462
\reference{} Bower, R., Lucey, J. \& Ellis, R. 1992, \mnras, 254, 589
\reference{} Bruzual, G. \& Charlot, S. 2003, \mnras, 344, 1000
\reference{} Butcher, H. \& Oemler, A., 1978, \apj, 226, 559
\reference{} Couch, W., Barger, A., Smail, I., Ellis, R. \& Sharples, R. 1998, \apj, 497, 188
\reference{} Couch, W., Balogh, M., Bower, R., Smail, I., Glazebrook, K. \& Taylor, M. 2001, \apj, 549, 820
\reference{} Dahlen, T., Fransson, C. \& Naslund, M. 2002, \mnras, 330, 167
\reference{} Davies, P. 2004, astro-ph/0403047 
\reference{} De Propris, R. and 29 co-authors 2004, \mnras, 351, 125
\reference{} Dressler, A. 1980, \apjs, 42, 565
\reference{} Dressler, A. \& Gunn, J. 1982, \apj, 263, 533
\reference{} Dressler, A., Oemler, A.,, Couch, W., Smail, I., Ellis, R., Barger, A., Butcher, H., Poggianti, B. \& Sharples, R. 1997, \apj, 490, 577
\reference{} Dressler, A., Smail, I., Poggianti, B., Butcher, H., Couch, W., Ellis, R. \& Oemler, A., 1999, \apjs, 122, 51
\reference{} Dressler, A. 2003, {\it Star Formation Through Time}, ASP Conference Proceedings, Vol. 297, p. 203
\reference{} Eggen, O., Lynden-Bell, D. \& Sandage, A. 1962, \apj, 136, 748
\reference{} Ellingson, E., Lin, H., Yee, H. \& Carlberg, R. 2001, \apj, 547, 609
\reference{} Ellis, R., Smail, I., Dressler, A., Couch, W., Oemler, A.,, Butcher, H. \& Sharples, R. 1997, \apj, 483, 582
\reference{} Faber, S. 1973, \apj, 179, 731
\reference{} Fasano, G., Poggianti, B., Couch, W., Bettoni, D., Kjaelig;rgaard, P. \& Moles, M. 2000, \apj, 542, 673
\reference{} Goto, T., Okamura, S., Yagi, M., Sheth, R., Bahcall, N., Zabel, S., Crouch, M., Sekiguchi, M.,
Annis, J. \& Bernardi, M. 2003, \pasj, 55, 739
\reference{} Gunn, J. \& Gott, J. 1972, \apj, 176, 1
\reference{} Larson, R. 1974, \mnras, 166, 585
\reference{} Jones, L., Smail, I. \& Couch, W. 2000, \apj, 528, 118
\reference{} Kauffmann, G. 1996, \mnras, 281, 487
\reference{} Kauffmann, G. \& Charlot, S. 1998, \mnras, 294, 705
\reference{} Kneib, J., Ellis, R., Smail, I., Couch, W. \& Sharples, R. 1996, \apj, 471, 643
\reference{} Kodama, T., Arimoto, N., Barger, A. \& Arag'on-Salamanca, A. 1998, \aap, 334, 99
\reference{} Kormendy, J. 1980, {\it Proc. ESO Workshop on Two-Dimensional Photometry}, p. 191
\reference{} Kuntschner, H. 2000, \mnras, 315, 184
\reference{} MacArthur, L., Courteau, S., Bell, E. \& Holtzman, J. 2004, \apjs, 152, 175
\reference{} Margoniner, V., d, Carvalho,, Gal, R. \& Djorgovski, S. 2001, \apj, 548, L143
\reference{} Moore, B., Katz, N., Dressler, A. \& Oemler, A. 1996, \mnras, 379, 613
\reference{} Odell, A., Schombert, J. \& Rakos, K. 2002, \aj, 124, 3061
\reference{} Pahre, M. 1999, \apjs, 124, 127
\reference{} Pimbblet, K. 2003, PASA, 20, 294
\reference{} Poggianti, B., Bridges, T., Carter, D., Mobasher, B., Doi, M., Kashikawa, N., Komiyama, Y., Okamura, S. \& Sekiguchi, M. 2001, \apj, 563, 118
\reference{} Poggianti, B., Smail, I., Dressler, A., Couch, W., Barger, A., Butcher, H., Ellis, R. \& Oemler, A., 1999, \apj, 518, 576
\reference{} Rakos, K., Maindl, T. \& Schombert, J. 1996, \apj, 466, 122
\reference{} Rakos, K. \& Schombert, J. 1995, \apj, 439, 47
\reference{} Rakos, K., Schombert, J., Odell, A. \& Steindling, S. 2000, \apj, 540, 715
\reference{} Rakos, K., Schombert, J., Maitzen, H., Prugovecki, S. \& Odell, A. 2001, \aj, 121, 1974
\reference{} Rakos, K. \& Schombert, J. 2004, \aj, 127, 1502
\reference{} Rakos, K. \& Schombert, J., 2005, \pasp, in press
\reference{} Schombert, J. 1986, \apjs, 60, 603
\reference{} Schombert, J. 1987, \apjs, 64, 643
\reference{} Schombert, J. 2005, in prep.
\reference{} Schulz, J., Fritze-v, Alvensleben,, Moumller, C. \& Fricke, K. 2002, \aap, 398, 89
\reference{} Smail, I., Kuntschner, H., Kodama, T., Smith, G., Packham, C., Fruchter, A. \& Hook, R. 2001, \mnras, 323, 839
\reference{} Smolcic, V., Ivezic, Z., Gacesa, M., Rakos, K., Pavlovski, K., Ilijic, S., Lupton, R., Schlegel, D., Kauffmann, G., Tremonti, C., Brinchmann, J., Charlot, S., Heckman, T., Knapp, G. \& Gunn, J. 2004, \mnras, in press
\reference{} Stanford, S., Eisenhardt, P. \& Dickinson, M. 1998, \apj, 492, 461
\reference{} Steindling, S., Brosch, N. \& Rakos, K. 2001, \apjs, 132, 19
\reference{} Trager, S., Faber, S., Worthey, G. \& Gonzalez, J. 2000, \aj, 120, 165
\reference{} van den Bergh, S. 1990, \apj, 348, 57
\reference{} van den Bergh, S. 1994, \aj, 107, 153
\reference{}  van Dokkum, P., Stanford, S., Holden, B., Eisenhardt, P., Dickinson, M. \& Elston, R. 2001, \aplett, 552, 101
\reference{} Visvanathan, N. \& Sandage, A. 1977, \apj, 216, 214
\reference{} Wang, D., Owen, F., Ledlow, M. \& Keel, W. 2004, astro-ph/0404313
\reference{} Ziegler, B., Bower, R., Smail, I., Davies, R. \& Lee, D. 2001, \mnras, 325, 1571

\end{references}
\end{document}